\documentclass[twocolumn]{emulateapj}
\usepackage{natbib}
\usepackage{amsmath}

\begin{document}
\title{Subluminous Type Ia Supernovae at High Redshift from the
  Supernova Legacy Survey$^{1}$}
\author{S.~Gonz\'{a}lez-Gait\'{a}n$^{2}$}
\author{K.~Perrett$^{2,3}$}
\author{M.~Sullivan$^{4}$}
\author{A.~Conley$^{2,5}$}
\author{D.~A.~Howell$^{6,7}$}
\author{R.~G.~Carlberg$^{2}$}
\author{P.~Astier$^{8}$}
\author{D.~Balam$^{9}$}
\author{C.~Balland$^{8}$}
\author{S.~Basa$^{10}$}
\author{D.~Fouchez$^{11}$}
\author{J.~Guy$^{8}$}
\author{D.~Hardin$^{8}$}
\author{I.~M.~Hook.$^{9,12}$}
\author{R.~Pain$^{8}$}
\author{C.~J.~Pritchet$^{13}$}
\author{N.~Regnault$^{8}$}
\author{J.~Rich$^{14}$}
\affiliation{$^{1}$Based on observations obtained with MegaPrime/MegaCam, a joint project of CFHT and CEA/DAPNIA, at the Canada-France-Hawaii (CFHT) which is operated by the National Research Council (NRC) of Canada, the Institut National des Sciences de l'Univers of the Centre National de la Recherche Scientifique (CNRS) of France, and the University of Hawaii. This work is based in part on data products at the Canadian Astronomy Data Centre as part of the Canada-France-Hawaii Telescope Legacy Survey, a collaborative project of NRC and CNRS.}
\affiliation{$^{2}$Department of Astronomy and Astrophysics, University of Toronto, 50 St. George Street, Toronto, ON, M5S 3H4, Canada}
\email{gonzalez@astro.utoronto.ca}
\affiliation{$^{3}$Network Information Operations, DRDC Ottawa, 3701 Carling Avenue, Ottawa, ON, K1A 0Z4, Canada}
\affiliation{$^{4}$Department of Physics (Astrophysics), University of Oxford, DWB, Keble Road, Oxford, OX1 3RH, UK}
\affiliation{$^{5}$Center for Astrophysics and Space Astronomy, University of Colorado, 593 UCB, Boulder, CO, 80309-0593, USA}
\affiliation{$^{6}$Las Cumbres Observatory Global Telescope Network, 6740 Cortona Dr., Suite 102, Goleta, CA 93117, USA}
\affiliation{$^{7}$Department of Physics, University of California, Santa Barbara, Broida Hall, Mail Code 9530, Santa Barbara, CA 93106-9530, USA}
\affiliation{$^{8}$LPNHE, Universit\'e Pierre et Marie Curie Paris 6, Universit\'e Paris Diderot Paris 7, CNRS-IN2P3, 4 Place Jussieu, 75252 Paris Cedex 05, France}
\affiliation{$^{9}$Dominion Astrophysical Observatory, Herzberg Institute of Astrophysics, 5071 West Saanich Road, Victoria, BC, V9E 2E7, Canada}
\affiliation{$^{10}$Laboratoire d'Astrophysique de Marseille, P\^ole de l'\'Etoile Site de Ch\^ateau-Gombert, 38, rue Fr\'ed\'eric Joliot-Curie, 13388 Marseille cedex 13, France}
\affiliation{$^{11}$CPPM, CNRS-IN2P3 and University Aix Marseille II, Case 907, 13288 Marseille cedex 9, France}
\affiliation{$^{12}$INAF, Osservatorio Astronomico di Roma, via Frascati 33, 00040 Monteporzio (RM), Italy}
\affiliation{$^{13}$Department of Physics \& Astronomy, University of Victoria, PO Box 3055, Stn CSC, Victoria, BC, V8W 3P6, Canada}
\affiliation{$^{14}$DSM/IRFU/SPP, CEA-Saclay, F-91191 Gif-sur-Yvette, France}

\begin{abstract}
  The rate evolution of subluminous Type Ia Supernovae is presented using data 
from the Supernova Legacy Survey. This sub-sample represents the faint and 
rapidly-declining light-curves of the observed supernova Ia (SN~Ia) population 
here defined by low stretch values ($s\leq0.8$). Up to redshift $z=0.6$, 
we find 18 photometrically-identified subluminous SNe~Ia, of which six have 
spectroscopic redshift (and three are spectroscopically-confirmed SNe~Ia). 
The evolution of the subluminous volumetric rate is constant or slightly decreasing 
with redshift, in contrast to the increasing SN~Ia rate found for the normal 
stretch population, although a rising behaviour is not conclusively ruled out. 
The subluminous sample is mainly found in early-type galaxies with little or no 
star formation, so that the rate evolution is consistent with a galactic mass 
dependent behavior: $r(z)=A\times M_g$, with $A=(1.1\pm0.3)\times10^{-14}\,
\mathrm{SNe}\,\mathrm{yr}^{-1}M_{\odot}^{-1}$.
\end{abstract}

\keywords{supernovae: general}


\section{Introduction}\label{intro}

While it is commonly agreed that Type Ia Supernovae (SNe~Ia) are thermonuclear
disruptions of mass accreting C-O white dwarfs (WDs) in a binary 
system \citep{Hoyle60}, the physics of the explosion and the nature of the
companion are still under discussion. The study of the properties and environments of SNe~Ia can help us
solve the progenitor question. Successful progenitor and explosion models need to explain the variety in light-curve shape and spectra of SNe~Ia. Recent multi-dimensional simulations of the explosion \citep{Gamezo05,Livne05,Kuhlen06,Kasen09} show the asymmetric character of the delayed detonation and succeed to explain the scatter in the width-luminosity relation of the normal SNe~Ia, although complications for the extreme SNe~Ia still exist. In particular, subluminous SNe~Ia -- a group of objects considerably fainter at peak (up to 2 mag), with faster light-curves, redder at early phases, with distinct spectral characteristics such as \ion{Ti}{2} and enhanced \ion{Si}{2} -- pose a challenge to any successful progenitor and explosion theory.

The prototypical example of a subluminous SN~Ia is SN1991bg \citep[e.g.][]{Filippenko92b} but many other examples have been discovered and studies of their properties relative to the bulk sample have been undertaken \citep{Garnavich04,Taubenberger08,Kasliwal08,Hicken09a}. Subluminous SNe~Ia are predominantly found in galaxies with older stellar populations, such as ellipticals and early-type spirals \citep{Howell01a}, in contrast to the preference for the late-type, star forming hosts favoured by bright, slow declining SNe~Ia \citep{Hamuy96,Hamuy00}. Subluminous SNe~Ia also occur exclusively in massive galaxies whereas the normal sample spans a wider host stellar mass range \citep{Neill10}. These observations set constraints on the delay-time -- the time between the formation of the binary system and subsequent SN explosion -- ranging from $< 1$ Gyr for SNe occurring in star-forming regions to several Gyrs for SNe in quiescent environments. They could as well hint at differences in the metallicity abundance of the progenitors, due to the mass-metallicity relation \citep{Tremonti04}, i.~e., subluminous SNe~Ia happen in more metal-rich environments.  

Based on a sample from the Lick Observatory Supernova Search (LOSS) and the Beijing Astronomical Observatory Supernova Survey (BAOSS), \citet{Li01,Li10b} found that $17.9^{+7.2}_{-6.2}\%$ of all local SNe~Ia are 1991bg-like, a possible overestimation as these surveys were host-targeted and could have systematically sampled brighter and more massive galaxies. The current observed subluminous sample remains a small fraction of the overall SN~Ia population, although new and recent surveys are actively looking for them \citep{Hicken09a}. At high redshift, they are challenging to identify spectroscopically, and the current high-$z$ SN surveys usually preferentially target normal SNe~Ia for cosmological purposes. As a result, no 91bg-like SNe have been located at $z>0.1$. For example, both, the Supernova Legacy Survey (SNLS) and the ESSENCE survey, report no spectra of 91bg-like objects at $z>0.1$, although this is consistent with their selection effects \citep{Bronder08,Foley08}.

This raises an obvious question: is this non detection simply due to the difficulty of detecting and classifying these objects, or is the relative frequency actually lower at high redshift because there has not been enough time for them to explode as SNe~Ia \citep{Howell01a}? Problems in the detection and classification arise from their intrinsic faintness, their rapidly evolving light-curves (which cause them to spend less time above the detection threshold for the same brightness), and their tendency to occur in brighter galaxies where the low contrast between the SN and host makes spectroscopy difficult. In this paper, we look for the fastest (and therefore faintest) SNe~Ia at $z>0.1$ in the Supernova Legacy Survey (SNLS) \citep{Astier06}. Even without spectroscopic follow-up, we can use the excellent multi-band light-curves of the SNLS to look for subluminous SN~Ia candidates by fitting subluminous SN~Ia LC templates to the photometric data.

Beyond a simple detection, even more enlightening would be a measurement of the evolution in the volumetric rate of these objects. As it is a convolution of the star-formation history and the delay-time distribution, the SN~Ia rate, in particular for sub-samples of SNe~Ia, can test different progenitor scenarios. Rates for SNe~Ia have been measured by several groups \citep[e.g.][]{Pain96,Pain02,Cappellaro99,Dahlen04,Dahlen08,Neill06,Dilday08,Dilday10,Perrett10,Li10c}. Combining the results one finds an increase with redshift up to $z\simeq1.0$, although with large spread among surveys. If the rate evolution for the individual SN~Ia populations differ, their delay-times must consequently vary and imply different conditions for the progenitors. As we probe higher redshifts, the SN~Ia environments will differ from the local sample, and will be reflected in the rates. If the fraction of high-mass (and high-metallicity) hosts was lower in the past, we would expect a similar behaviour for the rate of subluminous SNe~Ia.    



By constraining the delay-time distribution, we seek to improve our understanding of the progenitors of subluminous SNe~Ia. One can attempt to explain their progenitors in the SN~Ia progenitor framework, where two main scenarios have been envisaged: the single-degenerate (SD) model, in which a non-degenerate companion donates H/He-rich material to a WD near the Chandrasekhar mass \citep{Whelan73,Nomoto84}, and the double-degenerate scenario of two WDs coalescing \citep{Iben84,Webbink84}. In the SD scenario the delay-time is set by the age of the donor while in the DD model it depends on the age of the secondary and the merging time of the two WDs through gravitational wave radiation. The merging of two WDs has recently shown to be a viable mechanism, under certain conditions, for a successful subluminous explosion \citep{Pakmor10}. 

There is also a variety of independent mechanisms for subluminous explosions that explain the faintness by requiring a smaller burned mass of $^{56}$Ni. Instead of a common explosion model for the whole range of SNe~Ia, the delayed detonation \citep{Mazzali07} and other mechanisms like pure deflagrations or sub-Chandrasekhar CO or O-Ne-Mg WD off-center detonations \citep[e.g.][]{Livne90,Woosley94,Filippenko92b,Isern91} provide less luminosity although with some observational discrepancies \citep{Hoeflich96}. A failed neutron star model, in which C-O is accreted rapidly and ignited on the surface of the WD also leads to a faint SN~Ia \citep{Nomoto85}. Recently proposed models of WD collisions in dense environments could potentially lead to subluminous explosions \citep{Raskin09,Rosswog09}. Finally, a theoretically-motivated group of unusual objects, the so called SNe~.Ia (``dot Ia'') introduced by \citet{Bildsten07}, are the final helium shell flashes of AM CVn binaries. These are too dim and evolving too rapidly to be classical subluminous SNe~Ia, but possible candidates are starting to appear \citep{Poznanski10,Kasliwal10}.


With a detection of subluminous SNe~Ia at $z>0.1$, the measurement of the rate evolution and study of their host environments, we can start setting constraints on the evolution of these objects and their properties, their delay-times and progenitors. This paper is a companion to \citet{Perrett10}, who have carried out a rate study of normal SN~Ia sub-samples using the same SNLS data sample. We complement that study for the faint and short-lived SN~Ia population up to $z=0.6$. We use ``stretch'' as a LC-shape parameter \citep{Perlmutter97,Goldhaber01, Conley08} and define hereafter subluminous objects as SNe~Ia with $s\leq0.8$, inspired by the behavior of the color-stretch and magnitude-stretch relations (\S\ref{colstrels}). This value is higher than the ones inferred from other studies \citep[e.g.][]{Taubenberger08,Li10b}, as will be shown later. We emphasize therefore that our sample is less extreme and call it hereafter ``low-stretch'' (low-$s$) sample.

A brief outline of the paper follows. In \S\ref{lowz} we investigate the nearby training sample of low-$s$ SNe~Ia and find the necessary LC relations for creating a template. \S\ref{obs} highlights the observations and candidate selection of the SNLS. With this, we measure the rate as a function of redshift in \S\ref{rate_calc}, taking into account systematic effects from possible errors in the estimated LC parameters and contamination from core-collapse supernovae (CC-SNe). The discussion of the rate measurements together with the investigation of the host galaxy properties of the candidates follow in \S\ref{disc} and we conclude in \S\ref{conclusions}. We use a flat cosmology with $(\Omega_m,\Omega_{\Lambda})=(0.27,0.73)$ and $h=0.7$.


\section{Local relations for low-stretch SNe~Ia}\label{lowz}

As shown by \citet{Garnavich04,Taubenberger08,Kasliwal08} and \citet{Hicken09a}, subluminous SNe~Ia have clear photometric characteristics: red colors, a faint peak brightness, and a steeper relationship between LC shape and peak magnitude. In this section we derive new parameter relations based on a training sample of subluminous objects at low redshift. These relations will be of great importance to define proper low-$s$ SN~Ia LC templates and to correctly calculate the SNLS survey efficiency.

We use a local training sample from the literature (the low-$z$ Constitution set and low-$s$, Table~\ref{table_lows_objects}) that spans the stretch range, $0.45<s<1.30$, so that the behavior of low-$s$ SNe can be compared with the normal population. Adequate light-curve phase coverage is ensured by requiring observations in at least two filters between -15 and 8 and between 5 and 25 effective days (corrected for stretch and redshift) after maximum brightness in $B$-band. We also include a sample at higher redshift from the SNLS consisting of spectroscopically-confirmed SNe~Ia over $0.3<z<0.6$, which provides wider wavelength coverage compared to the low-$z$ sample.

\begin{figure}[htbp]
\centering
\includegraphics[width=0.85\linewidth]{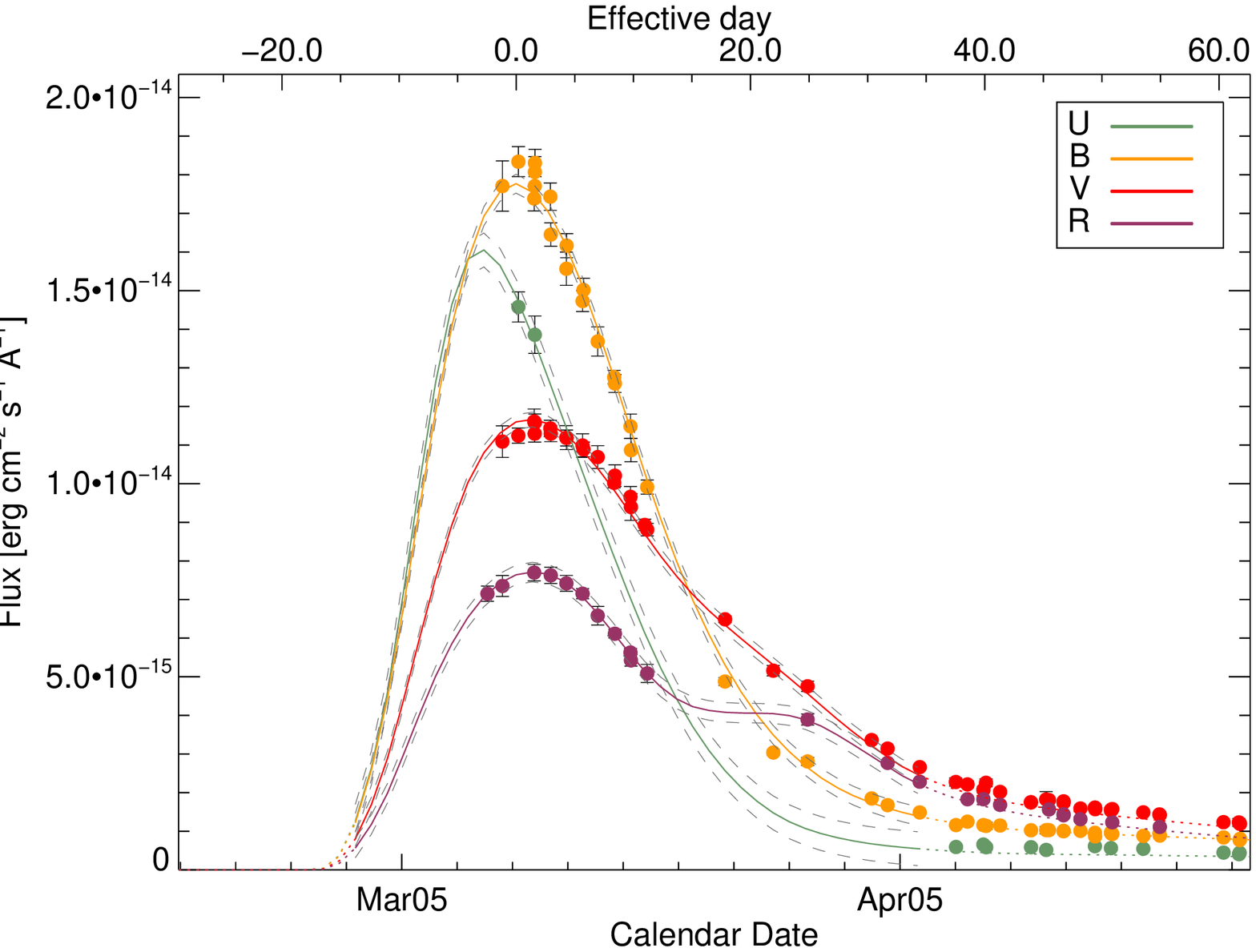}
\includegraphics[width=0.85\linewidth]{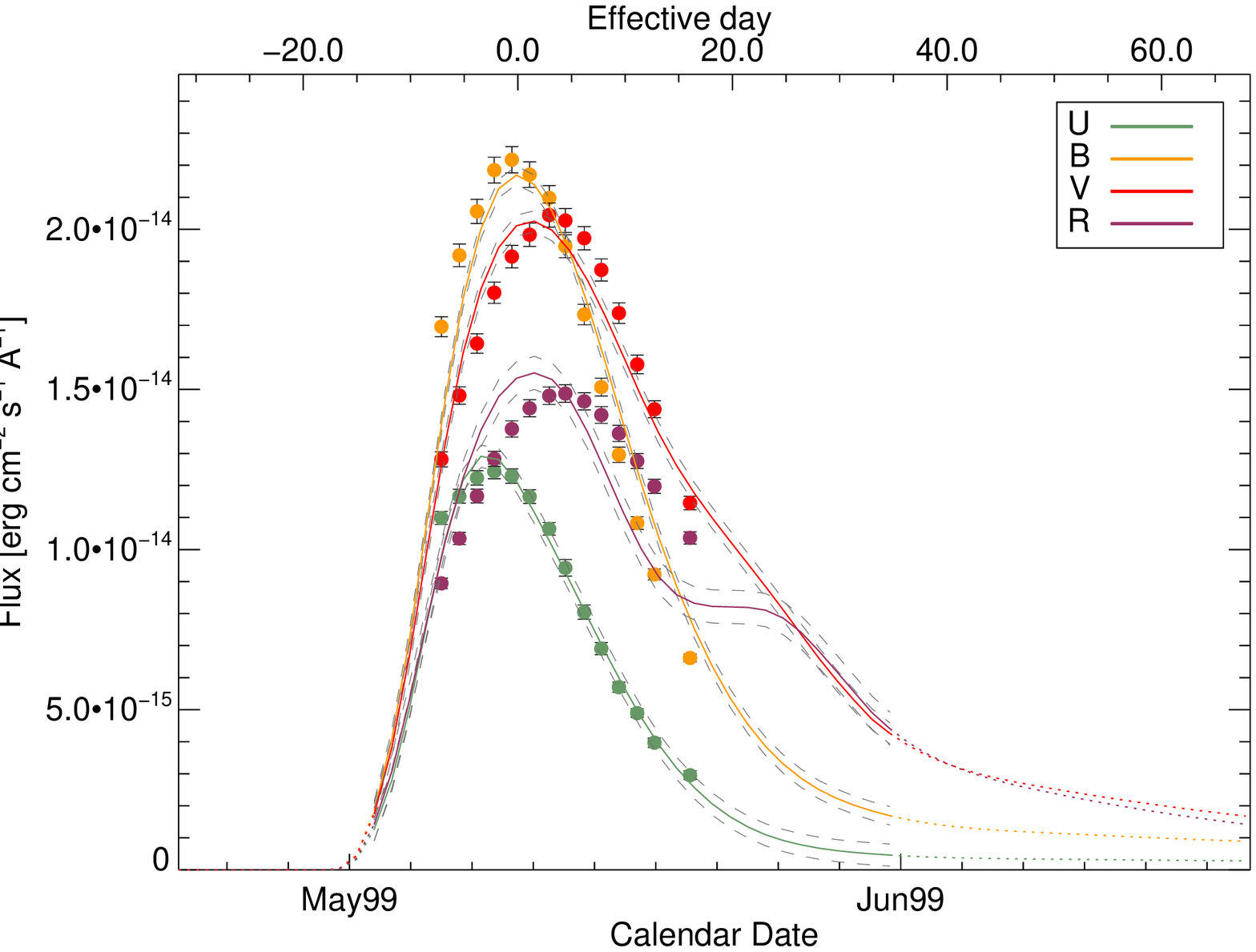}
\caption{Example SiFTO LC fits of low-$s$ SN2005am \citep{Hicken09a,Li06} with stretch 0.71 and overall $\chi^2_{\nu}=0.6$ and SN1999by (\emph{lower}) with stretch 0.61 and $\chi^2_{\nu}=8.5$. The fit is done for all points between -15 and 35 days in restframe.}
        \label{lowslowzfigs}
\end{figure}

We perform SN~Ia light curve fits to the entire sample with a modified version of the SiFTO algorithm \citep{Conley08}. SiFTO manipulates spectral energy distribution (SED) templates to model SN~Ia LCs. It has a wavelength-stretch dependence derived from strictly normal SNe~Ia, which was removed for the modified version used here. Although SiFTO was built with a training sample that excluded peculiar objects like SN1991bg, we obtain very good fits to subluminous data down to $s\simeq0.65$ and reasonable fits to lower stretch SNe (see Figure~\ref{lowslowzfigs}). Surprisingly, this template performs better than those built purely from subluminous objects like SN1991bg, which may have extreme features not suitable for all $s\leq0.8$ objects. The success of the fitting is partly because SiFTO fits each band independently without imposing any color-stretch relation, and so the unusual colors of subluminous SNe~Ia do not pose a problem. The algorithm provides the LC parameters such as stretch and peak magnitudes to get colors at maximum $B$-band brightness with their respective errors, as well as quality-of-fit parameters.

We use a reduced overall goodness-of-fit $\chi^2_{\nu}<5$ cut for normal and a higher $\chi^2_{\nu}<11$ for low-$s$ objects to account for the poorer fits of the more extreme objects. The final training set consists of: 207 low-$z$ SNe~Ia, of which 44 are low-$s$ (see Table~\ref{table_lows_objects}) with $s$ down to $s=0.45$, and 106 SNLS spectroscopic SNe~Ia of which 3 are low-$s$. A list of all culls is shown in Table~\ref{table_train_culls}. We note that not all of the low-$s$ SNe~Ia listed by \citet{Neill10} (Table 1) pass our selection criteria. Also, their stretch values for low-$s$ objects are in average lower than ours. This is due to the modified version of SiFTO we use, which assumes a common single stretch independent of wavelength and which has been more extensively tested for low-$s$ SNe~Ia. Additionally, despite our light curve coverage criteria, one low-$s$ candidate (SN2001da) does not have many data points and results in very different stretch ($s=0.89$) when fit through the method of \citet{Neill10}. We take this supernova out of our analysis. 

In order to develop tools to photometrically identify low-$s$ SNe~Ia and distinguish them from other transients, we also include a nearby and a high-$z$ SNLS sample of spectroscopically-identified non-type Ia supernovae. 14 nearby and 21 SNLS CC-SNe are properly fit to a SN~Ia SiFTO template, and 10 of those are assigned a $s\leq0.8$ stretch (6 of which are low-$z$, see Table~\ref{table_cont}). This set of contaminants will give us an idea of the systematic uncertainties (more in \S\ref{cont}).

Our limiting definition of low-$s$ objects at $s=0.8$ is inspired by the color-stretch and magnitude-color-stretch relations and is different from the ``subluminous'' definition studied in other surveys. We can compare it to the SALT2 LC-fitter \citep{Guy07} using similar equations to Eq. 6 in \citet{Guy10}, obtaining that $s=0.8$ for our modified SiFTO version corresponds to $X_1\simeq-1.85$. By directly matching subluminous SNe of \citet{Taubenberger08} and fitting a linear relation between their $\Delta m_{15}(B)$ parameter (the amount of $B$-band magnitudes the SN has declined 15 days after maximum) and our SiFTO stretch, we obtain a corresponding value of $\Delta m_{15}(B)\simeq 1.54$. This value is lower than the one they find to separate their sample ($\simeq1.70$, which would correspond to $s\simeq0.72$) and confirms again that our sample is not as extreme.

\subsection{Color-stretch relations}\label{colstrels}

Modeling the colors of SNe~Ia is fundamental for the LC fit and candidate selection of \S\ref{obs}, as well as for the study of the recovery efficiency needed for the rate calculation of \S\ref{rate_calc}. The challenge in modeling SN~Ia colors resides in the difficulty of disentangling intrinsic and extrinsic colors, the latter coming mainly from reddening of the host galaxy. As explained in \citet{Conley08}, the SiFTO method does not impose a single color model across all filters, which would probably not encompass the whole range of SN~Ia variability; rather the SED is adjusted to match the observed colors in each filter.

\begin{figure}[htbp]
\centering
\includegraphics[width=1.\linewidth]{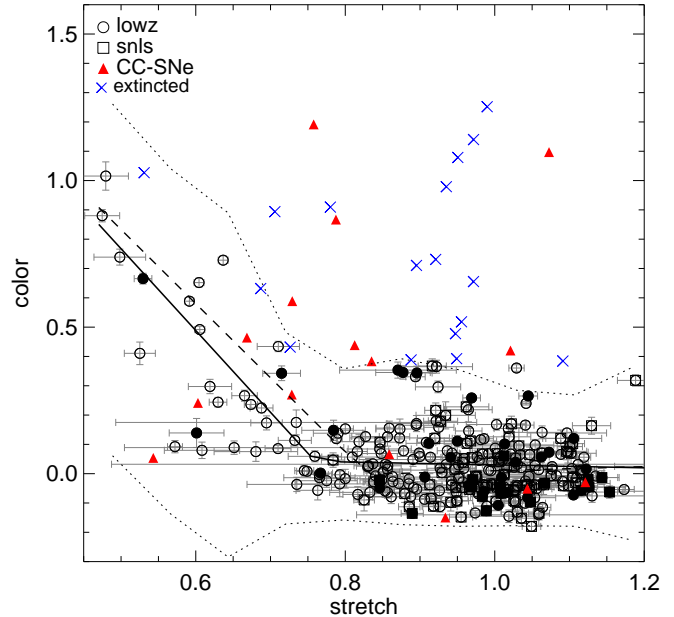}
\caption{Color as a function of stretch for the local SN~Ia sample (circles) and $0.3<z<0.6$ SNLS sample (squares). The filled points represent SNe in passive galaxies and the fit to exclusively those objects is the dashed line, a two-piece linear model changing at $s\simeq 0.8$. The ``low-extinction''  sample (open and filled black points) represent all objects with $c<0.37$ regardless of the host galaxy. The solid line is the best two-piece linear fit to this sample and the dotted lines are the the $2\times2\sigma$ error-snakes. High-extinction SNe (blue crosses) are above the maximum possible intrinsic colors calculated from the SNe in passive galaxies and were not included in the fits. CC-SNe (red triangles) are also shown.}
\label{color_stretch}
\end{figure}

We first remove highly reddened objects by the foreground Milky Way using the \citet{Schlegel98} dust maps and a cut of $E(B-V)_{\mathrm{MW}}<0.15$. To get an idea of the intrinsic color-stretch relation of the objects, we take the transients that are hosted in less extincted passive galaxies, i.e. in galaxies catalogued as E/S0 for the nearby sample from the NED database\footnote{The NASA/IPAC Extragalactic Database (NED) is operated by the Jet Propulsion Laboratory, California Institute of Technology, under contract with the National Aeronautics and Space Administration} and in galaxies with no star formation for the SNLS \citep{Sullivan06b}. A color similar to $B-V$ at B-band maximum brightness, $c$, as a function of stretch for these objects is shown in Figure~\ref{color_stretch} (filled circles and squares). 

We create a ``low-extinction sample'' by taking all SNe, regardless of their hosts, with a color below the upper limit from the maximum found for the SNe in only quiescent environments, $c<0.37$. This is only done for normal-$s$ SNe~Ia, i.~e. $s>0.8$, which have a relatively flat color-stretch relation. Low-stretch SNe~Ia become redder at a steeper rate with decreasing stretch and are hosted in less dusty environments, so we do not use such a maximum color cut, although we remove known extremely extincted objects (e.g. SN1986G, SN2006br and SN2007ax). The fit to this larger sample (open and filled circles and squares) in Figure~\ref{color_stretch} is well represented by a piecewise linear model with a break at $s=0.77$ (solid line):

\begin{equation}\label{colsteq}
c(s)=
\begin{cases}
C_0+\gamma_1\,(s-1) & (s>0.77)\\ 
C_0+\gamma_2\,(s-1) + C_{\mathrm{match}} & (s\leq0.77) \\ 
\end{cases}
\end{equation}  


The fit parameters are $C_0=0.08\pm0.02$, $\gamma_1=-0.06\pm0.02$, $\gamma_2=-2.79\pm0.07$  and the transition stretch value $s=0.77\pm0.01$.  $C_{\mathrm{match}}=-0.23(\gamma_1-\gamma_2)$ is a correction factor to unite both lines at $s=0.77$. The transition value for stretch ($s=0.77$), together with the one found for the magnitude-color-stretch relation in next section ($s=0.82$), argues for our chosen limiting value of $s\simeq0.8$ between low-$s$ and normal-$s$. The piecewise linear model has the best fit (reduced $\chi^2_{\nu}$ of 23.4) among other models (e.~g. exponential with $\chi^2_{\nu}=25.0$ or power law with $\chi^2_{\nu}=25.9$). The figure shows that low-stretch SNe~Ia are not only redder but their color-stretch relation is much steeper than for normal SNe~Ia. Errors in stretch and color were propagated in the fits. Allowing a fit to all objects, also extincted ones, provides consistent results (with a transition stretch at $s=0.83$) and is also shown in Figure~\ref{color_stretch}. It is worth to mention that if we force a piecewise linear model with a fixed lower transition value of $s=0.7$, more similar to characteristic ``subluminous'' surveys \citep{Taubenberger08}, the fit is slightly worse ($\chi^2_{\nu}=24.6$) and the slopes steeper. Due to the large scatter of colors at low-$s$, a range of consistent models (with $\chi^2<\chi^2_{min}+2$) is obtained for transition values of $0.69-0.87$.


\subsection{Luminosity-color-stretch}

We next derive a relation between the absolute peak magnitude as a function of stretch and color. To limit the effects of peculiar velocities, we require SNe in our training sample to have either $z_{\mathrm{CMB}}>0.01$ (CMB frame redshift) or a known distance to the host taken from the NED database.

\begin{figure}[htbp]
\centering
\includegraphics[width=0.9\linewidth]{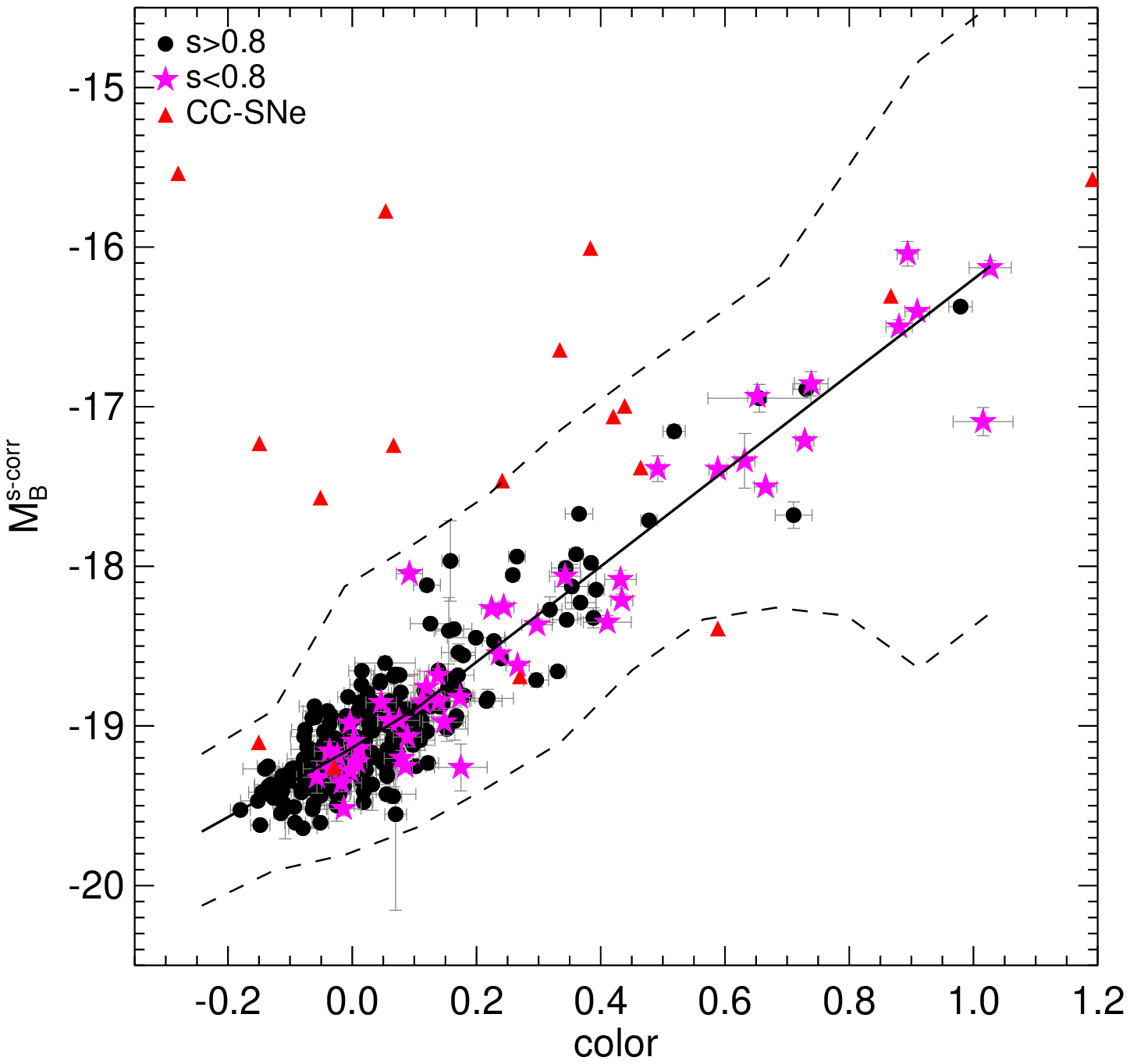}
\includegraphics[width=0.9\linewidth]{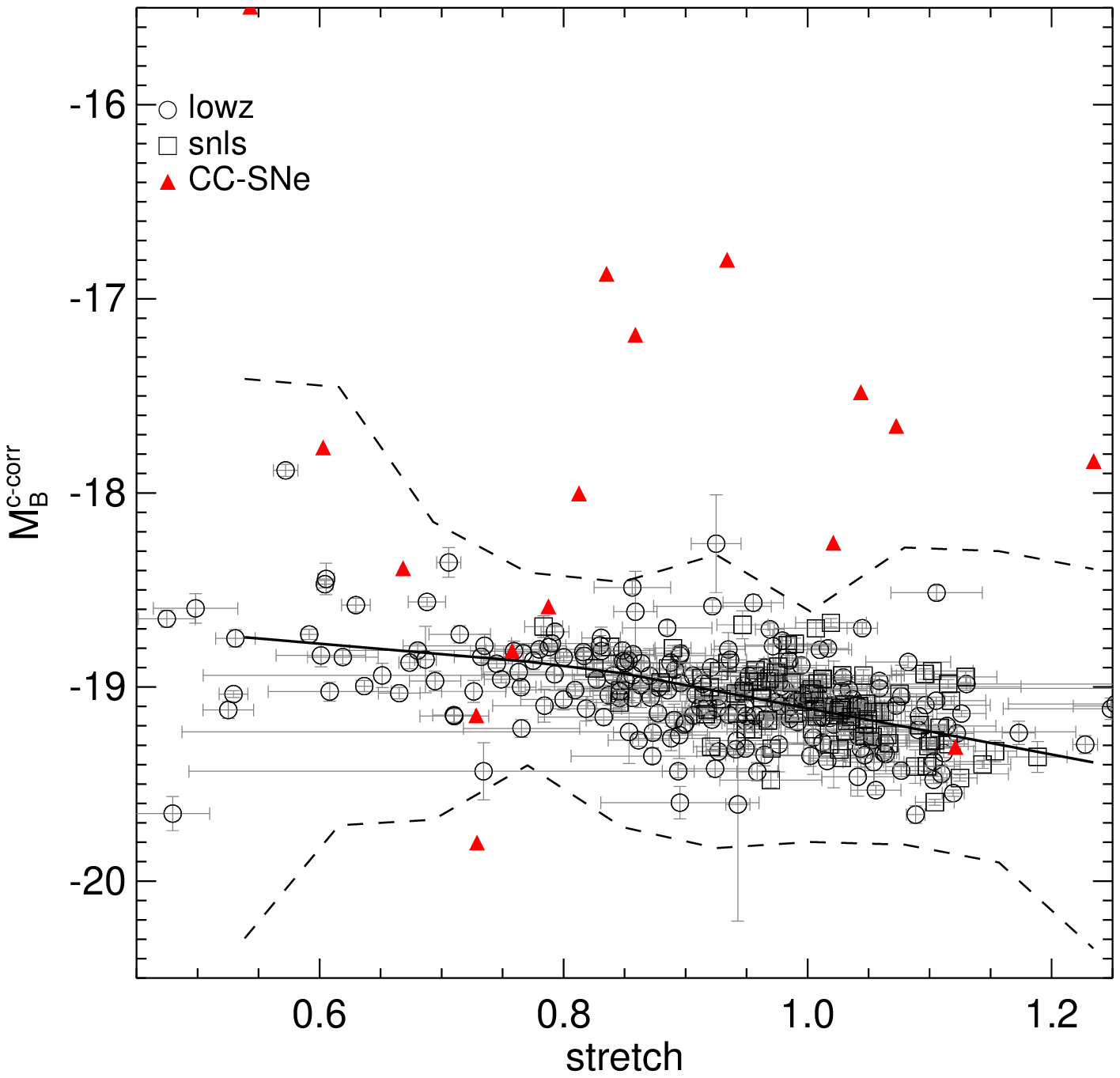}
\caption{Absolute magnitude (corrected for stretch from Eq.~\ref{magcolsteq}, i.~e. $M_B^{s-\mathrm{corr}}=M_B-\alpha_{[1,2]}(s-1)-M_2^{match}$) vs color (\emph{upper}) and absolute magnitude (corrected for color, i.~e.  $M_B^{c-\mathrm{corr}}=M_B-\beta_{[a,b]}c-M_b^{match}$) vs stretch (\emph{lower}) for the training sample. The solid lines represent the best fit model, a two-piece linear model for each case. The dashed lines are respectively the $2\times2\sigma$ error-snakes. CC-SNe (red triangles) are also shown.}
        \label{mag_color_stretch}
\end{figure}

We fit simultaneously magnitude, stretch and color to different models. The best one is again bilinear in both, stretch and color. The absolute magnitude shows a steeper slope as a function of color for redder objects changing at $c\simeq0.07$ (upper panel, Figure~\ref{mag_color_stretch}). Against stretch, the magnitude values  (lower panel, Figure~\ref{mag_color_stretch}) are also well fit by a bilinear model with a shallower slope for the low-$s$ stretch range. The effects in the differing slopes are not as strong as for the color-stretch relation but they will make the posterior analysis more precise. The errors in the absolute magnitude (from the apparent magnitude and redshift errors) and in the stretch and color are included in the fit. The magnitude-stretch-color relation can be summarized as follows:

\begin{equation}\label{magcolsteq}
M_B=M+\alpha_{[1,2]}\,(s-1) + \beta_{[a,b]}\,c + M_{[2,b]}^{match}
\end{equation}  

with $M=-19.11\pm0.01$, and where $\alpha_{1}=-1.19\pm0.03$ and $\alpha_2=-0.54\pm0.09$ are respectively used for stretch values greater and lower than the transition value of $s=0.82\pm0.02$.  $\beta_a=3.00\pm0.02$ and $\beta_b=2.15\pm0.05$ are used for values of $c$ greater and lower than $0.07\pm0.01$, respectively. $M_{2}^{match}=-0.18(\alpha_1-\alpha_2)$ and $M_b^{match}=0.07(\beta_a-\beta_b)$ unite the fit at the changing values. This model results in a slightly better fit ($\chi^2_{\nu}=19.6$) than other models (e.~g. exponential --in stretch and color-- with $\chi^2_{\nu}=19.9$).

We include an asymmetric error-snake on these relations, shown by the dotted lines in Figures~\ref{color_stretch} and~\ref{mag_color_stretch}. These delimit (twice) the region within which $95\%$ of all SNe~Ia in each bin of $0.1$ in color and $0.07$ in stretch are found around the best model. The width of the region has been doubled to allow for the larger photometric scatter of the high-$z$ SNe, which will be important later in \S\ref{finalcands}. This region also covers any uncertainties in the relations modeled in this section and allows for scatter in the color of the SNe. Requiring all objects to be inside those error-snakes, i.e. color-stretch, magnitude-color and magnitude-stretch, we find that $\simeq90\%$ of the normal and low-$s$ SNe~Ia pass the cut (see Table~\ref{table_train_culls}). However, only 2 contaminants (all SNe Ib/c) survive the same cull, a $\simeq6\%$ of the total of contaminants. We find this tradeoff to be beneficial. Reducing the contamination is useful when limiting the systematic errors as discussed in \S\ref{cont}.

\section{High-$z$ low-stretch candidate selection}\label{obs}

The search component of the SNLS, carried out using the Canada-France-Hawaii Telescope (CFHT), is a wide field imaging survey well-suited to the detection of SNe~Ia in the redshift range $0.1\leq z \lesssim1.0$ \citep{Astier06}. It is a rolling search where four $1^{\circ}\times1^{\circ}$ fields (D1 to D4) are imaged every few days, so that a well sampled LC is obtained in four filters ($g_Mr_Mi_Mz_M$) for each discovered supernova. Good candidates are followed-up spectroscopically. Because they are faint, less frequent, and not vital for cosmological studies, typical subluminous SNe~Ia are not prioritized in the spectroscopic follow-up \citep{Sullivan06a,Perrett10b}. We can still use the well-sampled SNLS multi-wavelength LCs to photometrically identify these objects. The method used resembles the one for the normal-stretch population of \citet[][hereafter P10]{Perrett10}, although with appropriate adjustments for the low-$s$ objects. This makes a direct comparison with the different stretch populations possible.

During the first 4 years of operation, up to July 2007, the candidate database consists of more than 5000 photometric transients including AGN, non-SNe and CC-SNe. As in P10, candidates in masked areas of the deep stacks are removed, as well as objects that lack the necessary LC coverage for a good photometric identification and parameterization. These observational criteria reduce the sample to half and are the same ones used for the SNLS training sample in \S\ref{lowz}: at least one $g_M$ observation between $-15$ and $5$ days of maximum, at least one $r_M$ and one $i_M$ observations between $-15$ and $+2.5$ days of maximum, at least one $r_M$ and one $i_M$ observations between $-9$ and $+7$ days of maximum, and at least one $r_M$ or one $i_M$ observation between $+5$ and $+20$ days of maximum.

\subsection{Redshift estimation and LC fit}\label{estsnz}

In P10, accurate photometric redshifts from the SNe were obtained with the use of two normal SN~Ia LC fitters. Their approach requires a good spectroscopic sample for testing the method and the precision of the final photometric redshifts. We do not possess any 91bg-like SN~Ia spectrum at high redshift with clear \ion{Ti}{2} features, only 3 low-$s$ objects at the border of our defintion, $s\lesssim0.8$, that have spectra of rather normal SNe~Ia. The high degree of degeneracy in the LC fits, particularly with stretch and redshift, which compete to make an object appear faint, demand an unbiased redshift estimate that will lead to appropriate LC parameters like stretch and color. 

Several hundred objects of the 4-year dataset possess spectroscopic redshifts (spec-z) from either the SN or from the host galaxy. For the SN-derived redshifts, the SN type has been determined from the spectra in the majority of cases \citep{Howell05,Bronder08,Ellis08,Balland09}. Most of these are normal SNe~Ia (413), although a good sample of CC objects is also available (78). The host spec-$z$ include data from DEEP/DEEP2 \citep{Davis03}, VVDS \citep{LeFevre05}, zCOSMOS \citep{Lilly07} and VIMOS \citep{Jonsson10}. Many candidates without spectroscopic redshifts have host photometric redshifts from the SNLS broad-band photometry either using the spectral connectivity analysis method of \citet{Freeman09}, or, when no match is found, through fits to spectral energy distributions as in \citet{Sullivan06b}. The resulting host photo-$z$, $z_{\mathrm{hostphot}}$, are very good to $z\simeq0.75$ with $\frac{|\Delta z|}{(1+z_{spec})}=0.042$ and a low ($\sim5\%$) catastrophic failure rate ($\frac{|z_{\mathrm{spec}}-z_{\mathrm{hostphot}}|}{(1+z_{\mathrm{spec}})}>0.15$).

Fortunately, some of those SNe~Ia with catastrophic redshifts can be identified with the help of the two LC fitters of P10 adjusted for low-$s$ SN: \texttt{estimate\_sn} \citep{Sullivan06a} and the aforementioned modified version of SiFTO (see \S\ref{lowz}). Their use in the multi-step process explained below provides good quality-of-fit parameters capable of distinguishing SNe~Ia from CC-SNe, but also offer an independent mechanism for obtaining photometric redshifts using the SN, $z_{\mathrm{SNphot}}$, in cases where host redshifts are lacking.
  
\begin{figure}[htbp]
\centering
\includegraphics[width=0.9\linewidth]{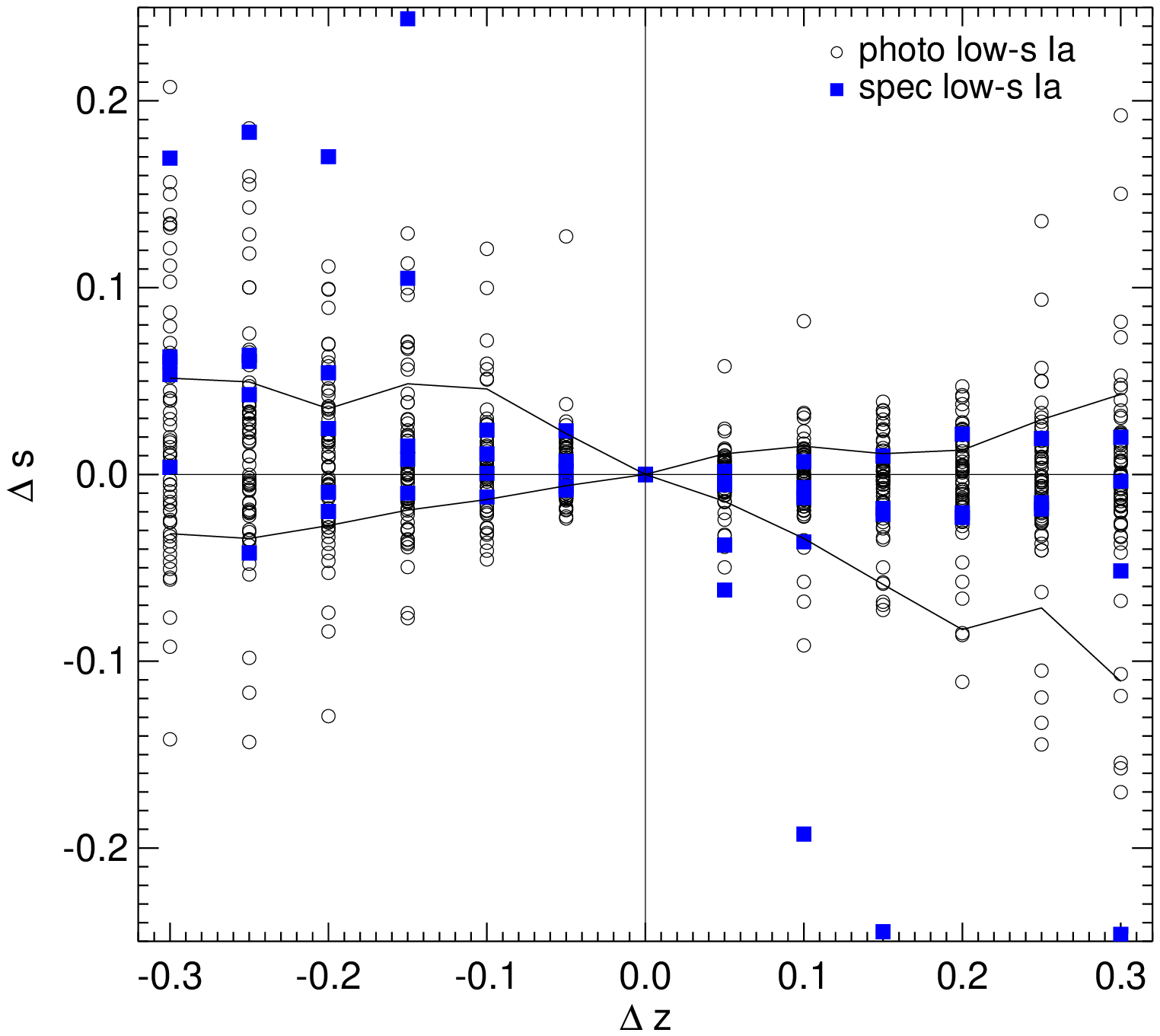}
\includegraphics[width=0.9\linewidth]{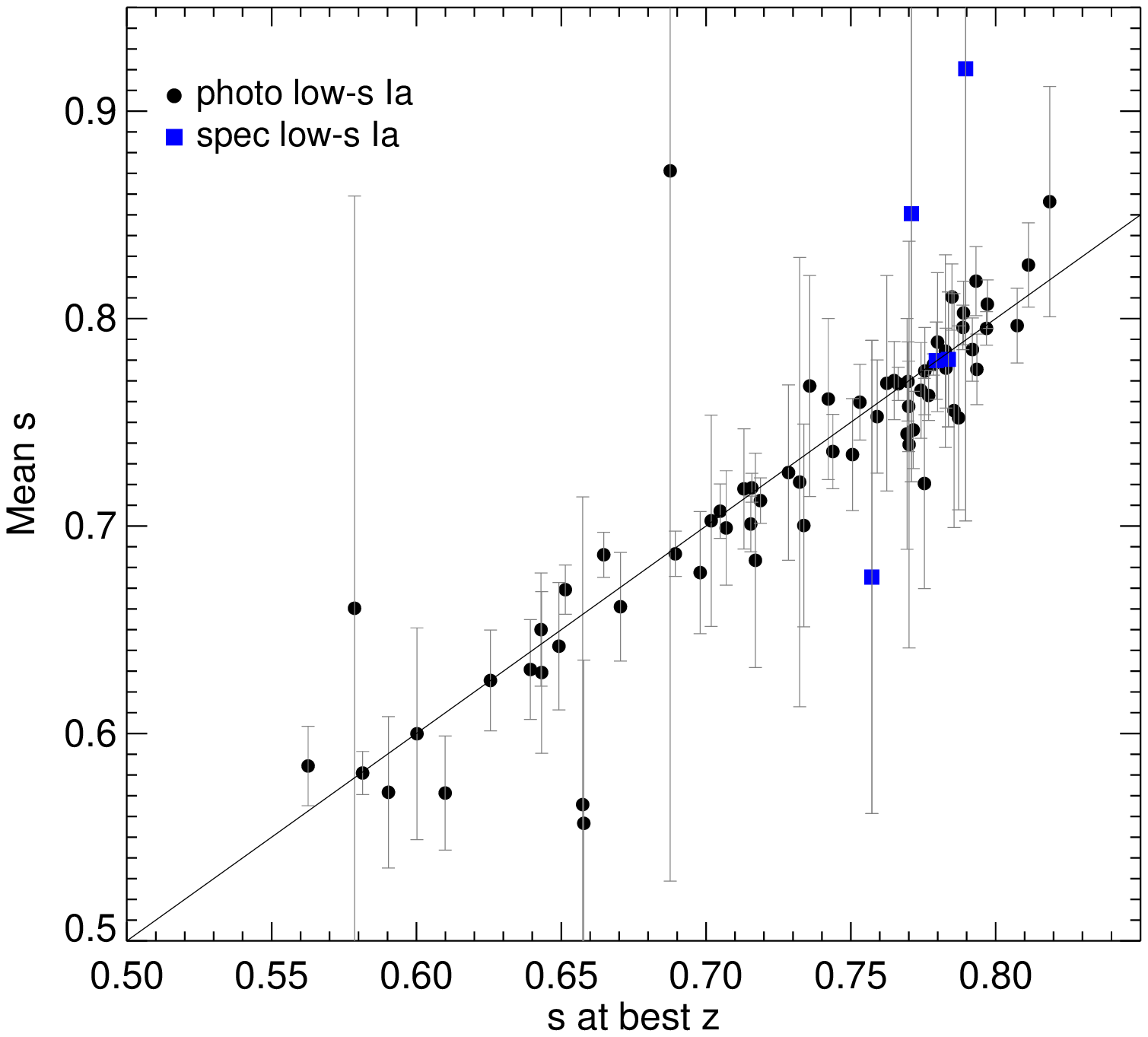}
\caption{\emph{Upper:} The change in fit stretch due to the deliberate variation in redshift for the low-$s$ final sample of \S\ref{finalcands}. The blue squares represent the spectroscopically-confirmed low-$s$ SNe~Ia. The solid black lines are the standard deviation above the median $\Delta s$. \emph{Lower:} The average output stretch for each SN~Ia as a function of its stretch (at zero redshift offset). The error bars are the change due to the variation in redshift for each SN~Ia.}
\label{s-variation}
\end{figure}

Once the redshift is known, SiFTO is used to obtain the stretch and color of each SN. Tests on the local sample (\S\ref{lowz}) show that it works well without imposing any stretch-color model. As shown in Figure~2 of P10, the derived stretch is largely independent of the redshift because the rise and fall of SNe~Ia are faster at shorter wavelengths, which largely compensates for redshift uncertainties. In Figure~\ref{s-variation}, we demonstrate that this behavior extends to low-$s$ objects. Shifting the redshift --either photometric or spectroscopic-- for each SN in our final low-$s$ sample (see \S\ref{finalcands}) by as much as $\Delta z=\pm0.3$, we obtain a stretch variation lower than $\sim s\pm0.1$. While this shows more dependence on redshift than the normal-stretch population, it still allows us to obtain accurate stretch estimates even in the presence of moderate redshift uncertainties.

The \texttt{estimate\_sn} routine is described in more detail in \citet{Sullivan06a}. It performs a grid search on the $\chi^2$ of a fit to SN photometry in flux space to a five-parameter model, including redshift. For this analysis, we modified it to include low-$s$ objects using the color-stretch relations of Eq.~\ref{colsteq}, and the magnitude relation of Eq.~\ref{magcolsteq}. This fit depends on the assumed values of the cosmological parameters.

We calculated the SN photo-$z$ for all objects in a similar fashion to P10 using the following multi-step process: First, we use \texttt{estimate\_sn} to get a preliminary redshift estimate. We then use this redshift as a fixed input to a SiFTO fit to obtain an improved stretch measurement. Finally, we feed back this stretch to a final \texttt{estimate\_sn} fit as a fixed parameter to obtain a more accurate $z_{\mathrm{SNphot}}$. We use this SN photo-$z$ for SNe that do not have any other redshift (spectroscopic or host photometric). 

Objects with $z_{\mathrm{hostphot}}$  that have ``bad fits'', i.e. that do not pass all cuts of \S\ref{finalcands}, are not discarded, but instead the cuts are re-applied using $z_{\mathrm{SNphot}}$. If this also fails, then the object is rejected. This offers some protection against catastrophic host photo-$z$ failures. This recovers 29 confirmed SNe~Ia. All known contaminants fail with both photo-$z$ values, so we are confident that this procedure still rejects non-Ias.

\subsection{Final candidates}\label{finalcands}

We perform final SiFTO (and low-$s$-modified \texttt{estimate\_sn}) fits to each object to obtain the parameters of each SN and quality-of-fit values. A series of cuts are then applied to screen out objects with poorly estimated parameters or those unlikely to be SNe~Ia. A summary of the applied culls and the number of objects surviving each of them can be found in Table~\ref{table_culls}. The $\chi^2$ culls are similar to those of P10, although we weaken the requirements slightly for $s\leq0.8$ objects based on the study of the low-$z$ sample: reduced $\chi^2$s of $\chi^2_{\mathrm{estimate\_sn}}<13$ for the overall fit, and for individual passbands $\chi^2_{\mathrm{estimate\_sn}}(r_M)<11$ and $\chi^2_{\mathrm{estimate\_sn}}(i_M)<16$, and a SiFTO cull of $\chi^2_{\mathrm{SiFTO}}<11$. Additionally, we require the objects to be inside the error-snakes trained in \S\ref{lowz} with low-$z$ low-$s$ SNe in order to constrain their parameters (Figure~\ref{new_color_stretch}). These cut values are derived empirically by trying to minimize the number of known contaminants that pass and minimize the number of rejected confirmed SNe~Ia.

\begin{figure}[htbp]
\centering
\includegraphics[width=1.05\linewidth]{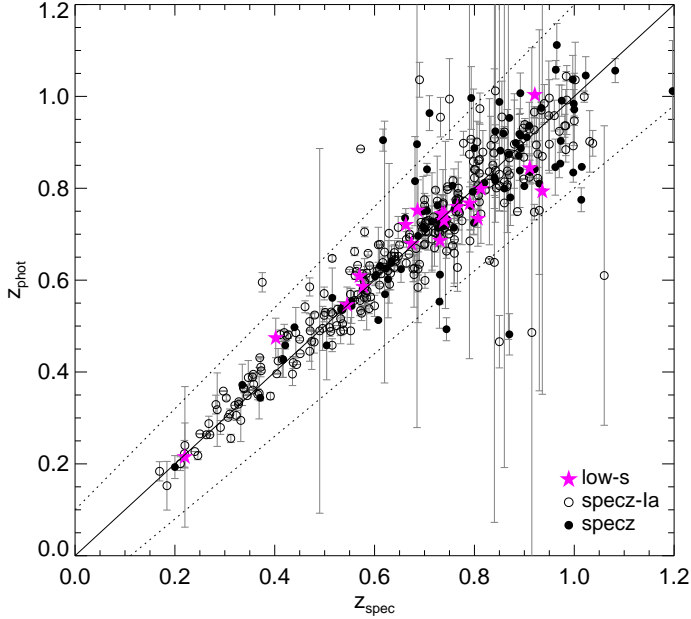}
\caption{Photometric redshifts obtained from the host or from the LC fits compared to spectroscopic redshifts from the SN (open circles) or the host (filled circles). Purple stars represent $s\leq0.8$ objects. The solid line shows 1:1 agreement and dashed lines a 10\% uncertainty in $(1+z_{\mathrm{spec}})$. The photo-$z$ errors are either from the host or from \texttt{estimate\_sn}.} \label{photz-specz}
\end{figure}

\subsubsection{At $z\leq0.6$}

Up to $z=0.6$, the final sample contains 161 objects, of which 126 have spectroscopic redshifts. They span the stretch range $0.6<s<1.2$, of which 18 have $s\leq0.8$. Six of these have known spectroscopic redshifts: three are confirmed SNe~Ia with $s\sim0.8$ with no outstanding 91bg-like feature in their spectra, and the other three have spectroscopic redshifts but have too low of a contrast with the underlying host galaxy to conclusively establish the spectroscopic SN type. The final parameter distributions are shown in Figure~\ref{finaldists} and Figure~\ref{finaldmag}. The low-$s$ photometric sample has median stretch and color of $\left<s\right>=0.74$ and $\left< c\right>=0.24$ (as compared to $\left<s\right>=1.00$ and $\left<c\right>=0.00$ for the normal-$s$ population). The mean $\left<\Delta\mathrm{mag}\right>=0.76\pm0.69$ clearly shows that our low-$s$ sample is not as extreme as the characteristic 91bg of $\Delta \mathrm{mag}\sim 2$mag. 

The complete final set of photo-$z$ contains 82\% with host photo-$z$ and 18\% with SN photo-$z$. Figure~\ref{photz-specz} shows the comparison of photometric and spectroscopic redshifts for all SN~Ia candidates with spectroscopic redshift either from the SN or the host. The median precision is $\frac{|\Delta z|}{(1+z_{spec})}=0.030$. The photo-$z$ are more accurate to $z\simeq0.6$ ($|\Delta z|/(1+z_{spec})=0.025$), where our rate study will concentrate.


\begin{figure*}[htbp]
\centering
\includegraphics[width=2.34in]{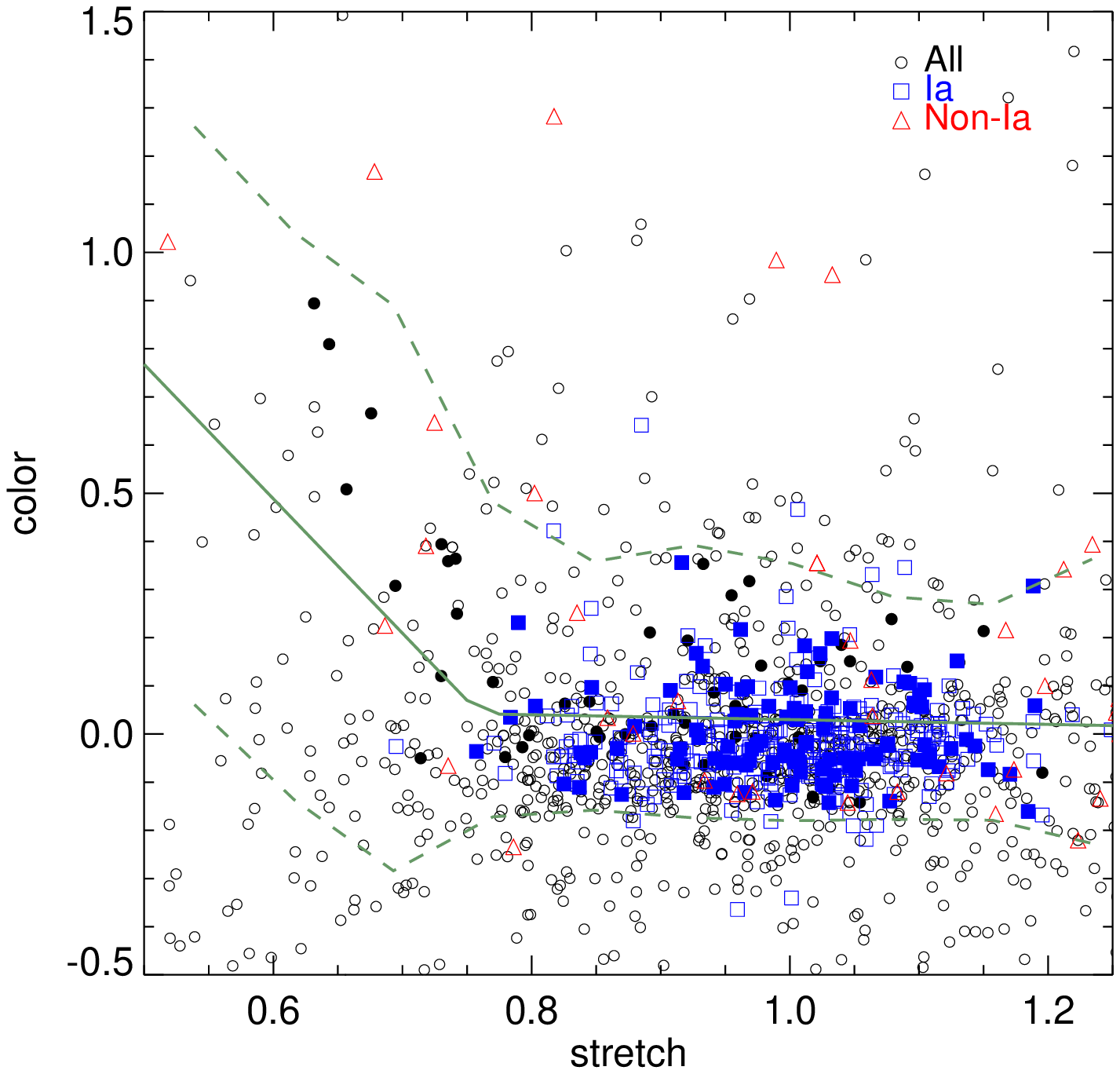}
\includegraphics[width=2.34in]{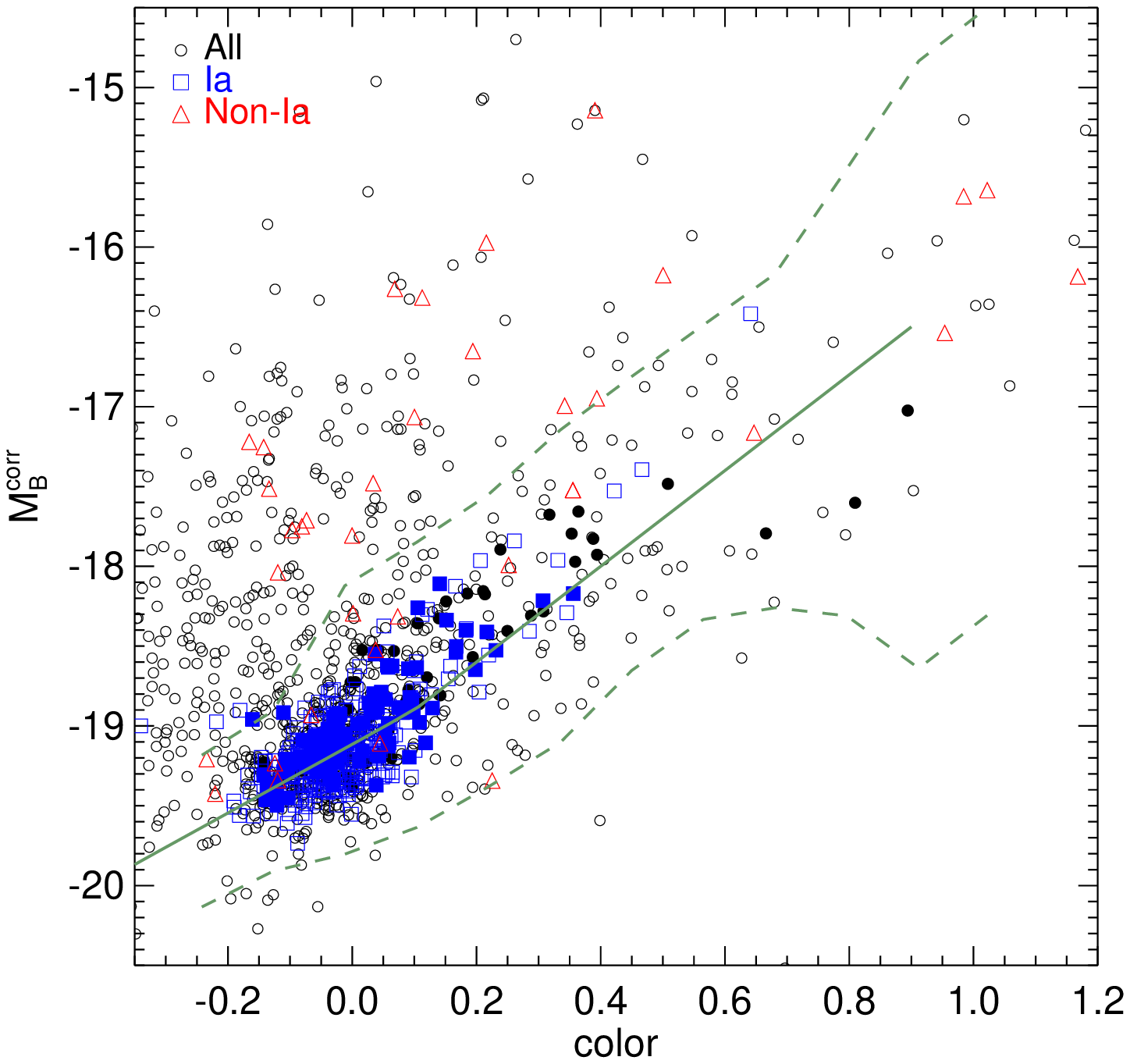}
\includegraphics[width=2.34in]{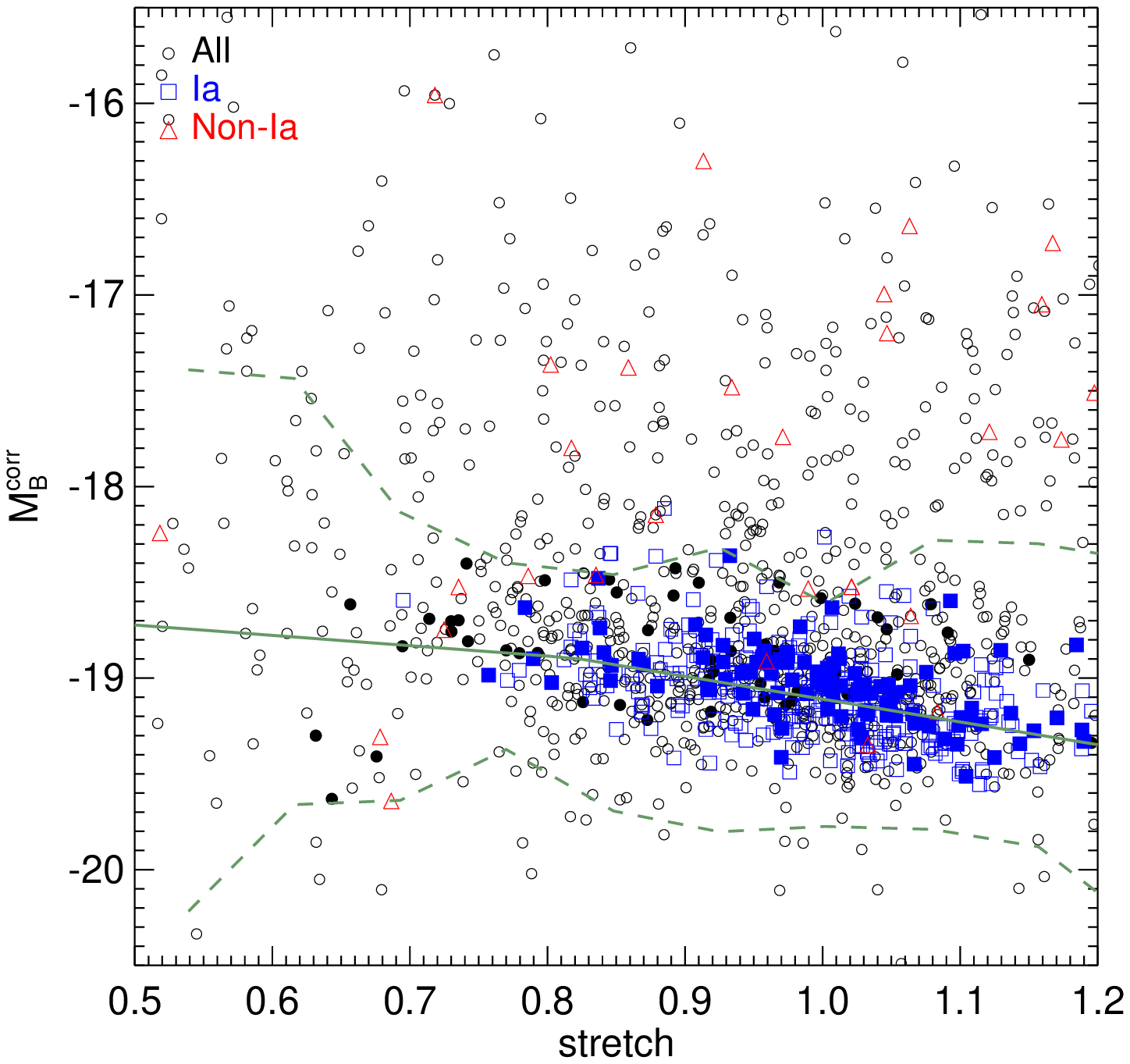}
\caption{Color-stretch (\emph{left}), magnitude-color (\emph{middle}) and magnitude-stretch (\emph{right}) for all SNLS objects passing the observation criteria (open points) and the final SN~Ia photometric $z<0.6$ sample after all cuts of Table~\ref{table_culls} (filled points). Spectroscopically-confirmed SNe~Ia (blue squares) and non-Ia (red triangles) are shown. The lines are the best-fit relations and error-snakes found in \S\ref{lowz} for the local training sample, and served as additional cuts.}\label{new_color_stretch}
\end{figure*}

The final errors in photo-$z$ are obtained from Figure~\ref{photz-specz} by taking the standard deviation around the median as a function of z. The stretch and color errors are provided by SiFTO, which requires a known redshift input. Stretch and color variations due to redshift uncertainties will be taken into account with a Monte Carlo simulation in \S~\ref{sims}.

\begin{figure*}[htbp]
\centering
\includegraphics[width=2.25in]{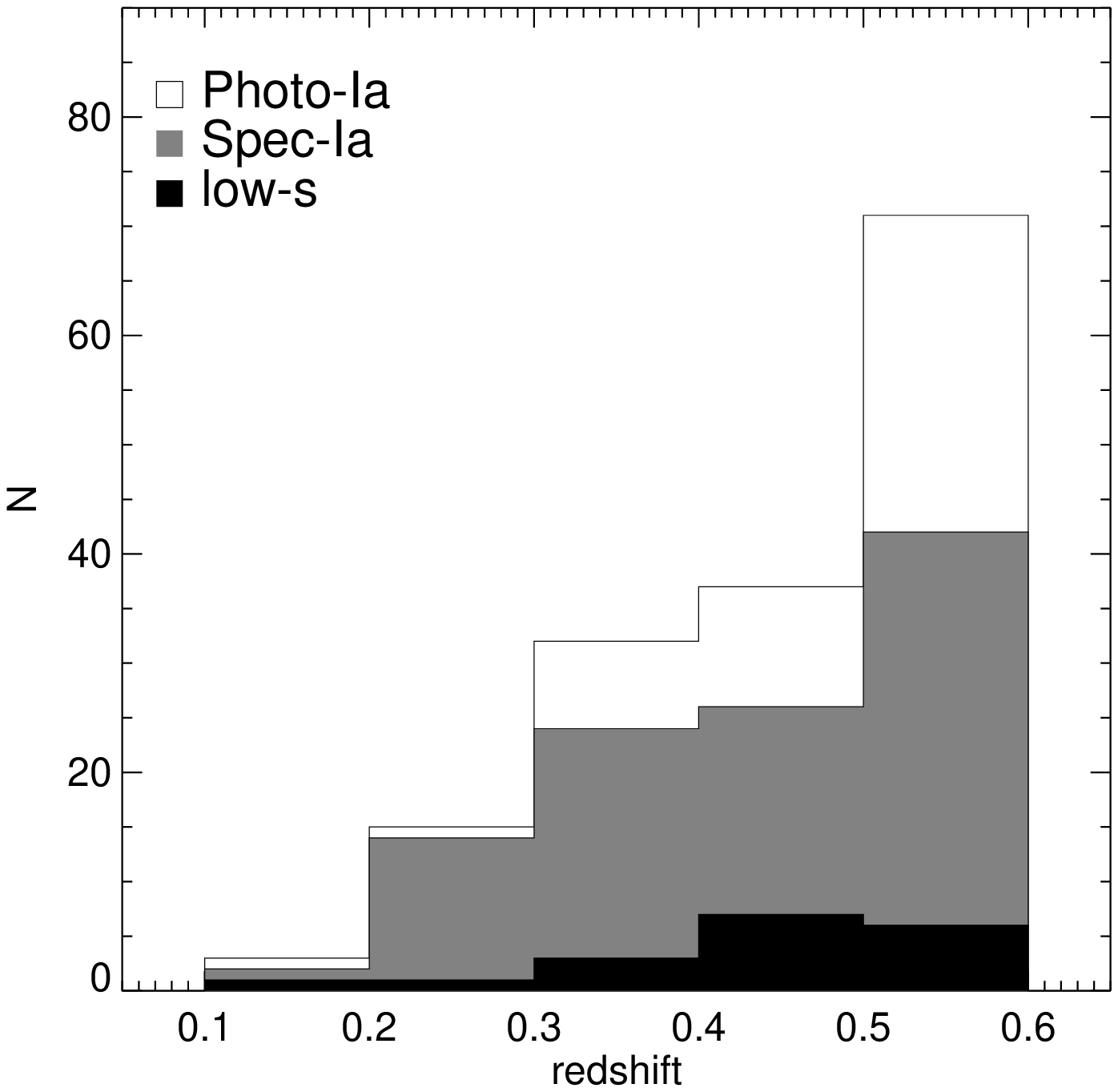}
\includegraphics[width=2.25in]{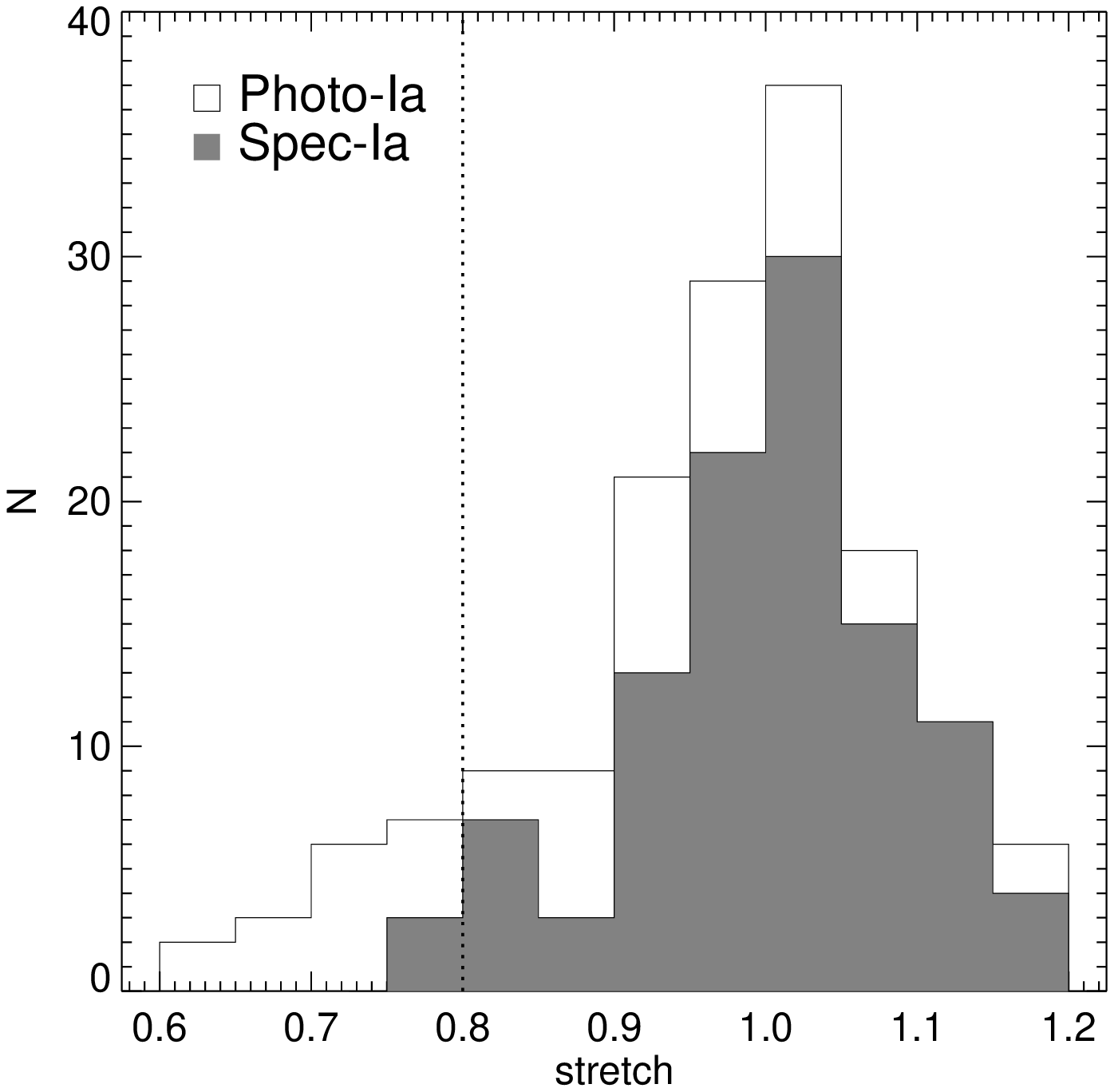}
\includegraphics[width=2.25in]{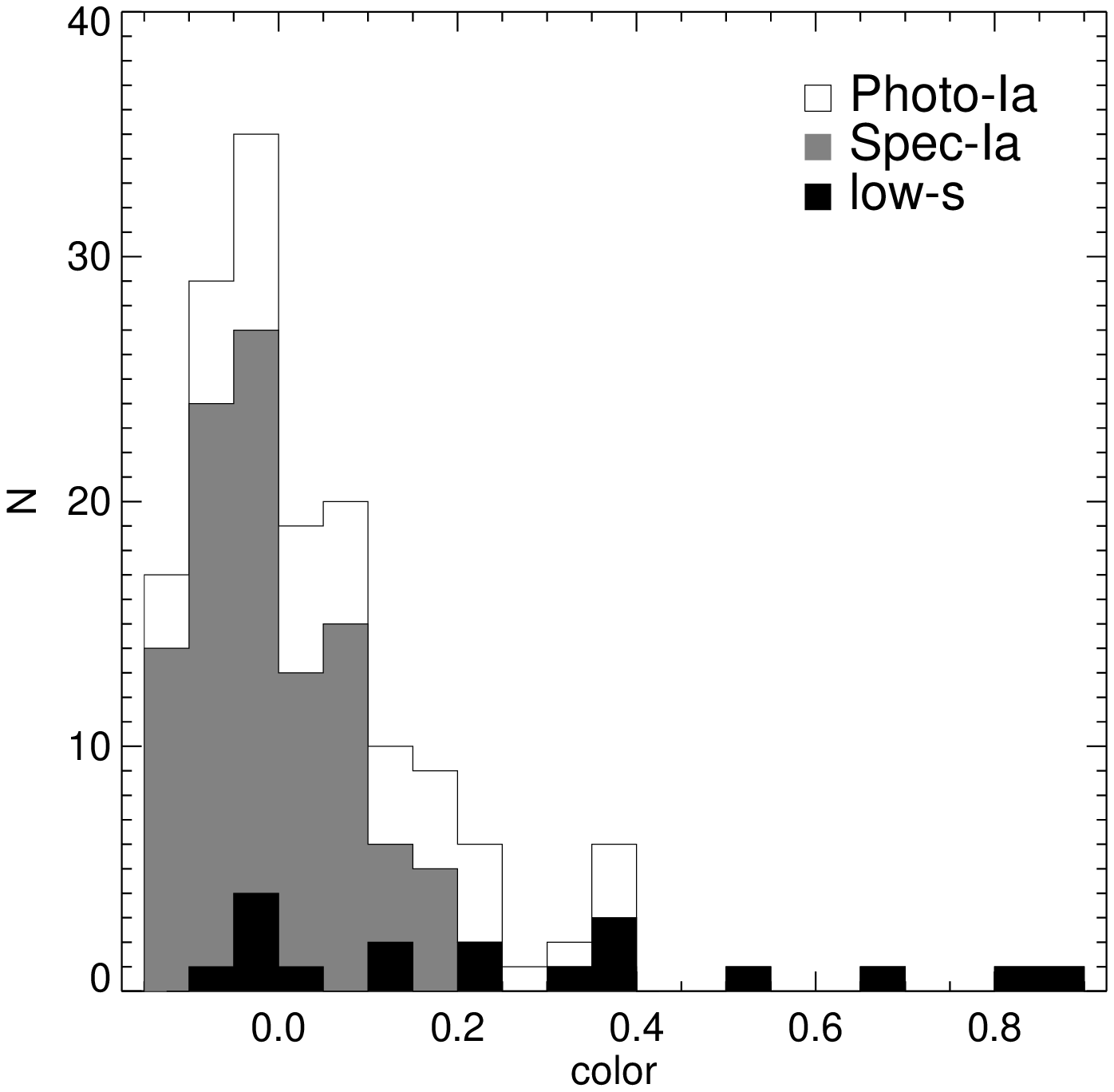}
\caption{Final distributions in redshift (left), stretch (center) and color (right) of the $z<0.6$ SN~Ia photometric sample including low-$s$. The white histogram is the whole population, the gray is the spectroscopically-confirmed SNe~Ia and the black the low-$s$ candidates.} \label{finaldists}
\end{figure*}

\begin{figure}[htbp]
\centering
\includegraphics[width=0.8\linewidth]{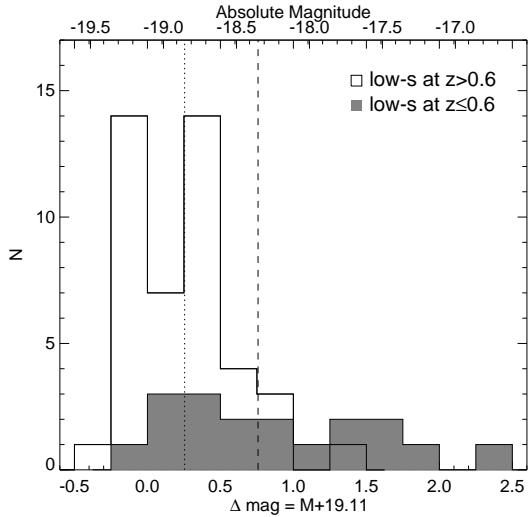}
\caption{Distribution in $\Delta \mathrm{mag}$ of the final low-$s$ candidate sample. $\Delta mag=M+19.11$ is the difference between the absolute magnitude of each SN compared to the magnitude $-19.11$mag of a ``standard"  SN~Ia at $s=1$ and $c=0$ according to eq.~\ref{magcolsteq}. The shaded histogram shows only the low-$s$ SNe~Ia at $z<0.6$.} \label{finaldmag}
\end{figure}

\subsubsection{At $z>0.6$}

\begin{figure}[htbp]
  \centering
  \includegraphics[width=0.82\linewidth]{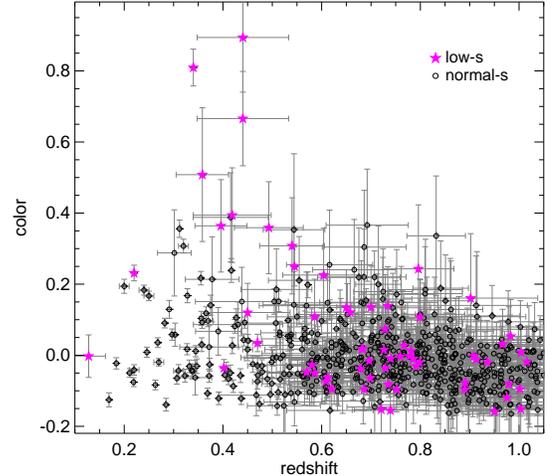}
  \caption{Color as a function of redshift for the final photometric SN~Ia sample. Normal-$s$ (black circles) and low-$s$ (purple stars) are highlighted.}
        \label{sz-cz}
\end{figure}

The detection incompleteness at $0.6<z<1.0$ increases rapidly for low-$s$ SNe~Ia, as will be shown in next section. We still find a significant population of 44 candidates (11 with spec-$z$ from the host and 5 with low signal-to-noise spec-$z$ from the SN, 2 of which are confirmed SNe~Ia of $s\sim0.8$ and no 91bg features) with median stretch, $\left<s\right>=0.76$, and color, $\left<c\right>=-0.02$. The average color is extremely blue compared to the median color at $z<0.6$ of 0.24 confirming that we are only observing the blue tail of the low-$s$ distribution. This can be seen in Figure~\ref{sz-cz}. They are not very faint objects, even for low-stretch, as shown in Figure~\ref{finaldmag}. The lack of characteristic low-$s$ red objects makes it difficult to reliably estimate the low-$s$ rate at $z>0.6$ and proper corrections need to be made (see \S~\ref{sims}).

Additionally, we do find a presence of blue low-$s$ candidates at high-$z$ which are absent at lower-$z$. Such blue objects with $c<-0.05$ are neither found in the nearby nor in the $z<0.6$ SNLS samples, but represent $\sim30\%$ (14) of all $z>0.6$ low-$s$ objects. We do not expect all of these objects to be CC-SNe, which could potentially contaminate the low-$s$ sample with their characteristic bluer colors. As will be shown later, besides going through our SN~Ia pipeline, the environment of these candidates is not characteristic of CC-SNe. 

A second possibility is that this group represents another class of transients that are fast and blue. The object found by \citet{Perets09} is faint and fast-evolving but with clear different spectroscopic features and bluer colors than 91bg-like objects. It is hosted in an old stellar population as well as most of our candidates. However, our objects are not as faint (Figure~\ref{finaldmag}). On the other hand, the transients found by \citet{Poznanski10} and \citet{Kasliwal10}, suggested SN~.Ia-like helium detonations of a WD, are blue and of similar brightness than ours, but evolve faster. When fit with our templates, we obtain very poor fits with $s=0.44$ and $s=0.33$, respectively, lower than our lower stretch limit for subluminous SNe~Ia. 

\begin{figure}[htbp]
\centering
\includegraphics[width=0.9\linewidth]{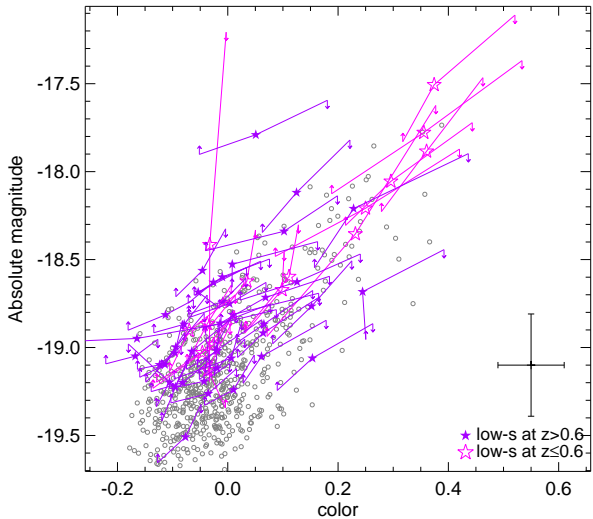}
\caption{Absolute magnitude vs color for the low-$s$ SN~Ia sample with photo-$z$ at $z<0.6$ (big open pink stars) and at $z>0.6$ (small filled purple stars). The dots show the normal-$s$ sample. The upward and downward arrows show the new magnitudes and colors when varying the photo-$z$ by $+0.05$ and $-0.05$ respectively. The lines unite the same object at the three redshifts. The lower right error bars represent the typical (mean) photometric errors.}
\label{mag-col-variation}
\end{figure}

A tempting explanation is that the input photometric redshift for the LC fit is consistently wrong and affects the obtained color and stretch. This hypothesis is ruled out because 7 of these objects have spectroscopic redshift and the rest (except one) have photometric redshift from the host (with mean photo-$z$ error of $\sim0.08$), which do not show any significant systematic. We have seen the small effect in the measured stretch when the redshift is varied (Figure~\ref{s-variation}). We now show the effect in the magnitude and color of photometric candidates in Figure~\ref{mag-col-variation}. Their variation due to changes in photo-$z$ cannot account for all the bluer colors but for some of them --only if the redshifts were all consistently overestimated by $+0.05$, the colors would be redder by $+0.08$ (and magnitudes fainter by $+0.17$, whereas stretches would only change by $-0.004$).

Photometric errors in stretch and color get larger as we probe higher redshifts. Most of the candidates found at these redshifts have stretch close to the limiting definition of $s=0.8$, so that they are consistent with being fast normal-$s$ SNe~Ia. Only 3 candidates have $s+s_{\mathrm{error}}<0.8$, one of them with $s<0.6$. Of these,two also have $c+c_{\mathrm{error}}<-0.05$, inconsistent with typical low-$s$ colors, and will be further discussed in ~\ref{cont}. The rest can therefore be explained as the blue tail of $s\sim0.8$ objects while the redder counterpart remains unseen due to selection effects. In the remainder of this work, we treat this group as such and correct for this selection bias when calculating the rates. 

\section{Low-stretch SN~Ia rate evolution}\label{rate_calc}
With a set of low-$s$ candidates, we proceed to calculate a volumetric rate as a function of redshift. As in P10, the volumetric rate, $r_b(z_b)$, for a given redshift bin, represented here by the middle of the redshift bin, $z_b$, is given by the sum over all $N_b$ candidates inside the bin:

\begin{equation}\label{calcrate}
r_b(z_b)=\frac{1}{V_b}\sum_{i=1}^{N_b}\frac{(1+z_i)}{\epsilon_i\Delta T_i},
\end{equation}

where $\epsilon$ is the individual SN detection efficiency depending on the characteristics of each object (like redshift, stretch and color) and $\Delta T$ is the sampling time accessible to find the object, divided by $(1+z)$ to account for time dilation. $V_b$ is the comoving volume contained in the redshift range of the bin.


\subsection{Detection efficiencies}\label{effs}
The detection efficiencies $\epsilon_i$ account for observing biases measuring the rate. They depend on numerous factors and are specified as a function of the SN characteristics such as redshift, stretch and color (along with field and year of observation). The efficiencies used here are described in P10 and \citet{Perrett10b}, and are based on a Monte-Carlo study of 2.4 million artificial SNe~Ia injected into the actual SNLS images. This approach avoids the need to make assumptions about temporal and spatial data coverage. 

The artifical SNe are generated using uniform distributions of redshift and stretch (down to $s=0.5$), and the color is determined using a relation like Eq.~\ref{colsteq}, including an intrinsic scatter around the stretch-color relation (a Gaussian noise with dispersion $\sigma=0.04$ added independently to $U$ and $V$). The peak magnitudes are generated from the luminosity-stretch-color relation, as in Eq.~\ref{magcolsteq}, and the redshift and assumed cosmology, plus an additional unmodeled magnitude dispersion from a Gaussian distribution with an intrinsic scatter of $\sigma_{int}=0.15$. The LC is generated from integrating, at a given epoch and through the filter functions, the SN Ia spectral templates scaled for the $B$-magnitude and corrected for $U-B$ and $B-V$ colors, Milky Way and host extinction. The artificial SN is then inserted into each $i_M$ image obtained by the SNLS. The images are then processed through the same pipeline used in the real-time SN discovery, and the recovered fraction is used to derive the efficiency as a function of various parameters, including redshift, stretch, color, observing season and field (see a more detailed description in \citealt{Perrett10b}). The low-$s$ candidates, which are faint, red and have a rapid rise and fall, are more difficult to detect and therefore have a lower efficiency, so are accordingly given a higher weight when calculating the rates. 

\begin{figure}[htbp]
  \centering
  \includegraphics[width=1.\linewidth]{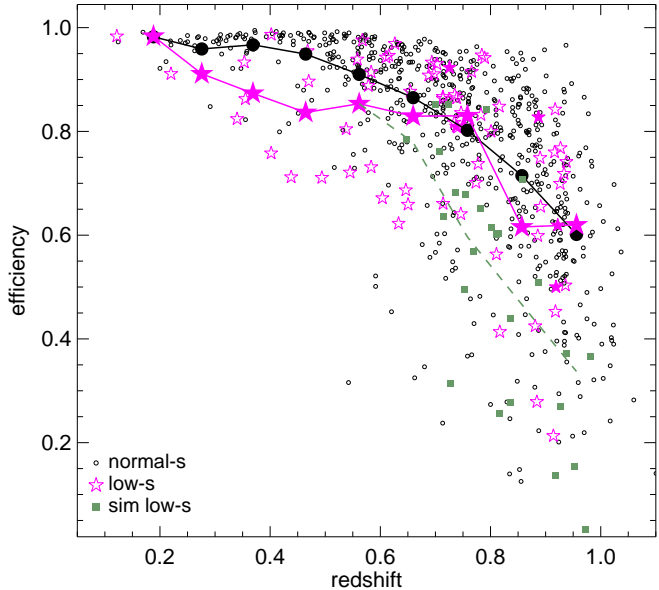}
  \caption{Efficiencies calculated by P10 for the SN~Ia photometric sample for normal-$s$ (black circles) and low-$s$ (purple stars) populations. For objects where no direct efficiency was found (small filled stars), the closest match in color was used. Mean values in redshift bins for normal-$s$ and low-$s$ populations (big filled circles and stars) are shown. Additionally, simulated low-$s$ objects for $z>0.6$ bins (green squares) based on the $z<0.6$ color distribution and respective mean (green dashed line) are shown for comparison. Note that these efficiencies do not include the observation criteria of section \S~\ref{obs} and sampling time correction of the different fields observed.}
        \label{effmeas}
\end{figure}

The efficiencies are calculated in bins of width $\Delta z=\Delta s=\Delta c=0.1$. The measured efficiencies, linearly interpolated according to each SN, are shown in Figure~\ref{effmeas}. It can be seen that the mean efficiencies are lower for the low-$s$ sample up to $z\sim0.6$. Above $z=0.6$, the mean efficiencies seem to approach the normal-$s$ sample. At these redshifts they become low for all but the bluest low-$s$ objects. This leads to a higher fraction of low-$s$ blue objects with very few red ones recovered, as found in \S~\ref{finalcands}. We show the expected efficiency distribution for the missing red, low-$s$ SNe~Ia of an unevolving color distribution as green squares in Figure~\ref{effmeas} based on the observed low-$s$ color distribution at low redshift from \S\ref{finaldists}. The rates at $z>0.6$ will need to be corrected for these missing objects (see next section). We note that the efficiencies shown in Figure~\ref{effmeas} are the raw efficiencies --the recovery fraction from the simulated SNe-- without the observational constraints of section \S~\ref{obs} nor the sampling time correction that depends on each field and season. These corrections can make the actual efficiencies much lower (see P10). For instance, one of our low-$s$ candidates at high-$z$ with $z=0.80$, $s=0.63$ and $c=0.24$ has a raw efficiency of 0.54 but with observational constraints and sampling time this efficiency is only 0.18, which explains why these objects are hard to observe.


\subsection{SN~Ia Rates and Monte-Carlo simulations}\label{sims}
\begin{figure*}[htbp]
  \centering
  \includegraphics[width=1.0\linewidth]{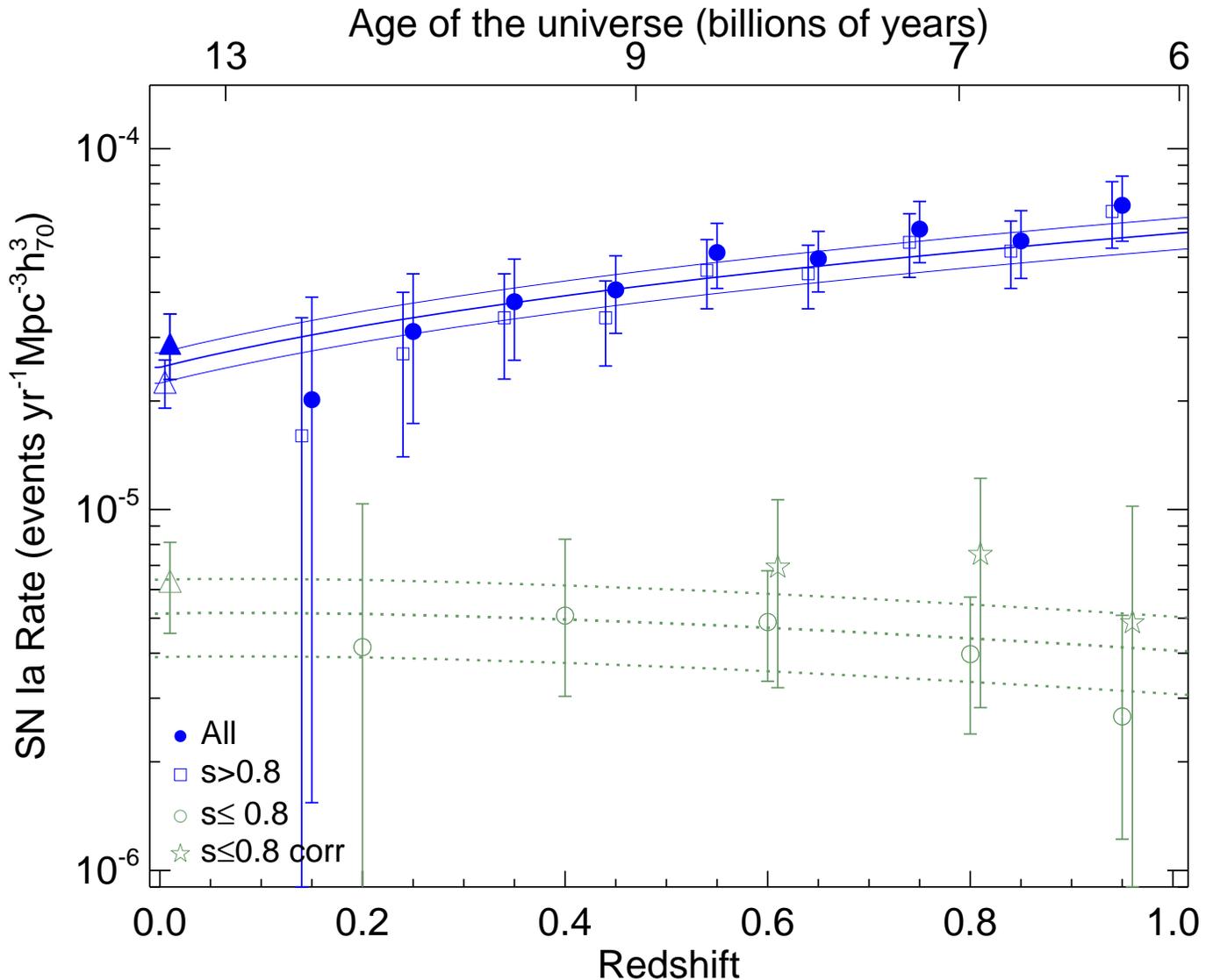}
    \caption{Volumetric rate of low-$s$ SNe~Ia as a function of redshift (green open points) compared to the normal-$s$ rate (blue squared points) of P10. The total summed rate (blue filled points) is shown with a little $z$-offset for clarity. The stars represent the corrected high-$z$ low-$s$ rate calculated from real and simulated objects drawn from a distribution scaled to the $z<0.6$ color distribution. The filled triangle is the total nearby SN~Ia rate from \citet{Li10c} and the open triangle is the low-$s$ rate estimated from a $s\leq0.8$ fraction of it. The solid lines are the best $A+B$ model and uncertainties found in P10. The dashed lines are the best A-model and uncertainties found in this study through a fit of only SNLS $z<0.6$ bins: $A=(1.05\pm0.28)\times10^{-14}\,\mathrm{SNe}\,\mathrm{yr}^{-1}M_{\odot}^{-1}$.}
        \label{rates}
\end{figure*}

We calculate a rate evolution, $r_{\mathrm{meas}}$, shown in Column 5 of Table~\ref{table_rates} solely based on Eq.~\ref{calcrate} and with the efficiencies from \S~\ref{effs}. We use bins of $\Delta z=0.2$ due to the low sample statistics (the last $z$-bin is only 0.1 wide as no low-$s$ candidate exists beyond $z=1$). The measured errors in stretch and color from the LC fit, and especially the errors in the photometric redshift of each candidate, can lead to uncertainties in the rate measurement. We account for these via a Monte-Carlo (MC) simulation that randomly shifts the photometric redshift according to its error (for those SNe~Ia with photo-$z$). For each redshift, a corresponding LC fit provides new parameters and its covariance matrix. This latter is used to generate a new random set of stretch and color using standard techniques \citep{James75}. We recalculate the rates from Eq.~\ref{calcrate} for each iteration based on the randomized values, and use the median value over the 1000 realizations to determine the final rate, $r_{\mathrm{MC}}$. We estimate the error from the variance of the rates in each bin (and call them ``MC" errors shown in Column 6 of Table~\ref{table_rates} and Figure \ref{rates}) and add it in quadrature to the weighted $1\sigma$ Poisson statistical errors in Figure~\ref{rates} (confidence levels calculated using the Clopper-Pearson method used for example in \citealt{Gehrels86}). The error simulation differs from P10, as the SNLS does not possess a known spectroscopic sample of low-$s$ SNe~Ia from which we could obtain parameter distributions. Instead, we treat each SN individually by randomly drawing parameters according to its errors and covariances. Although the precise details of our simulation differ from P10, a SN~Ia rate evolution within the errors is obtained.

\begin{figure}[htbp]
  \centering
  \includegraphics[width=0.8\linewidth]{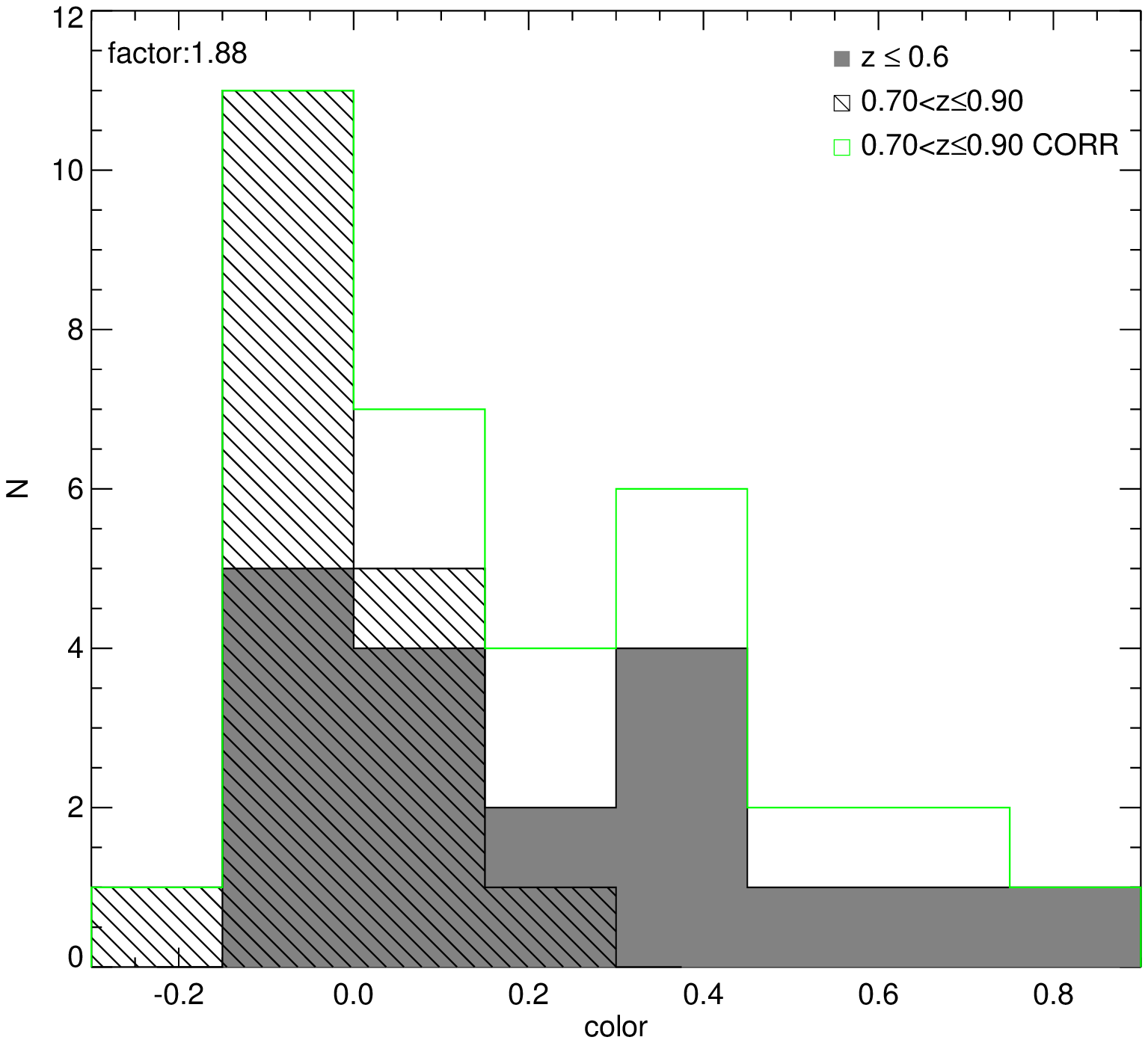}
 \caption{Example of how the observed color distribution for low-$s$ at $0.7<z<0.9$ (diagonal-shaded) is corrected to match the efficiency-corrected low-$s$ color distribution at $z<0.6$ (shaded). The final corrected color distribution for that $z$-bin (open histogram) has $f_b=1.88$ more SNe than the original observed one.}
        \label{highzdist}
\end{figure}

Additionally, we have seen in the previous section that we do not find red objects in the higher-$z$ bins. This selection bias results in underestimated rate values at high-$z$, which are not corrected by the efficiencies of the bluer objects found. There is a lack of objects in certain red color bins, so that no efficiency can account for them. To measure this effect, we take the low-$s$ distribution of color in an underestimated high-$z$ bin (for each MC iteration), we normalize this distribution to the known color distribution at $z<0.6$ (right Figure~\ref{finaldists}), which has already been corrected for the efficiencies of each SN, and then generate fake SNe in the missing color bins. The normalization is done with the two bluest matching bins. Figure~\ref{highzdist} shows an example for $0.7<z<0.9$. The green open histogram is the final corrected histogram from the initial observed diagonal-shaded distribution. The correcting factor for this case is $f_{b,\mathrm{corr}}=1+\frac{N_b(\mathrm{fake})}{N_b}=1.9$, where $N_b$(fake) is the number of generated SNe in the bin and $N_b$ is the observed number, so that the corrected rate is $r_{b,\mathrm{corr}}=r_b\times f_{b,\mathrm{corr}}$. The final correcting factors from the median of all MC iterationns, summed Poisson (from the number of objects used in the bins used to normalize both distributions) and MC errors, as well as the resulting corrected rates are shown in the last two columns of Table~\ref{table_rates}. In Figure~\ref{rates}, the corrected rates are shown as stars.

Finally, errors in the efficiencies can affect the rate values. As explained in P10, those come mainly from systematic errors in the underlying assumed $\Delta$mag distribution of SNe~Ia, the distribution of residuals between the absolute magnitude of a fiducial $s=1$ SN~Ia and the magnitudes of observed SNe~Ia after correction for stretch and color. To check for this systematic error, we allow for random efficiencies in the MC simulation calculated from a range of $\sigma_{int}=0.12-0.15$ in the $\Delta$mag distribution, obtaining results that agree very well with the rates of Figure~\ref{rates}.

\subsection{Contamination uncertainties}\label{cont}

An important systematic error comes from core-collapse SNe mimicking the LC of low-$s$ SNe~Ia. In P10, a missclassification fraction of less than $\leq3\%$ was found for the normal-$s$ population. We expect this fraction to be higher for low-$s$ SNe~Ia due to two effects. First, they are dimmer, and more easily confused with the intrinsically fainter CC-SNe at a given redshift. Second, as we probe higher redshift ranges, CC-SNe at low-$z$ might masquerade as low-$s$ SNe~Ia at higher-$z$ since we do not have spectroscopic redshifts for all candidates. SNe~II have characteristic LCs that are very different from SNe~Ia and present less danger of contamination. SNe~Ib/c, on the other hand, represent a smaller fraction of CC-SNe, $\sim30\%$ according to \citet{Cappellaro99,Smartt09} and $25\%$ according to \citet{Li10b}, but are more similar to SNe~Ia and will be the most probable contaminants.

The first error --where the photo-$z$ is correct, yet the fit of a SN~Ia template manages to pass the quality-of-fit cuts-- can be examined using the training sample (as in \S\ref{lowz}). We found a 6\% CC-SN missclassification rate, all SNe~Ib/c.  This fraction can be propagated to higher redshift using the SN~Ia rate evolution of P10, $R_{Ia}(z)\propto (1+z)^{\alpha_{Ia}}$ with $\alpha_{Ia}\simeq1.91$, and a CC-SN rate evolution reflecting the star formation rate from \citet{Hopkins06}, $R_{CC}\propto (1+z)^{\alpha_{CC}}$ with $\alpha_{CC}\simeq3.6$. We assume a ratio of SNe~Ia to CC-SNe of $\sim0.22$ at $z=0.3$ from \citet{Bazin09}, which translates to a local ratio of $\sim0.39$ -- as compared to $\sim0.48$ found by \citet{Cappellaro99} and $0.24$ from \citet{Li10b} --. The percentage fraction of contamination as a function of $z$, $f_{\mathrm{cont}}(z)$, can then be described as:

\begin{equation}
f_{\mathrm{cont}}(z)=\frac{N_{\mathrm{passcont}}(z)}{N_{\mathrm{pass}}(z)}=\frac{1}{1+0.39\, r\,(1+z)^{\alpha_{\mathrm{Ia}}-\alpha_{\mathrm{CC}}}},
\end{equation}

where $r=0.88/0.06$ is the cull passing efficiency (the probability of a SN passing all selection criteria) ratio of Ia with respect to CC, assuming it does not evolve with $z$. This equation allows us to get a crude estimate of the number of contaminants that would pass all cuts, $N_{\mathrm{passcont}}(z)$. By dividing the expected fraction of contaminants per $z$-bin by the mean weight of low-$s$ SNe found therein (per MC simulation), we translate this into a rate error lower than 20\% for $z<0.6$ up to as high as 30\% at $z\simeq0.9$. The estimates are sensitive to the power law of the rate evolution, the initial fraction of SNe~Ia to CC-SNe and the local missclassification fraction of $6\%$. Increasing this latter fraction to $10\%$ or using SN~Ia to CC-SN local ratio of $0.241^{+3.7}_{-3.8}$ by \citet{Li10b} instead, we obtain contamination percentages up to 30\% for $z<0.6$ and 40\% at $z\simeq0.9$. This latter rate error from contamination is added in Table~\ref{table_rates} and we believe it to be an upper limit. The contamination error is smaller than the combined statistical and MC errors at $z<0.6$ but becomes comparable at higher-$z$. One should note that the missclassification error increases with redshift and would only make the rate higher than the actual one.

The second source of contamination comes from uncertainties in the photo-$z$. Incorrect redshift estimates can cause low-$z$ CC-SNe to simulate low-$s$ SNe~Ia at high-$z$. This effect can be tested by running all spectroscopic objects through our selection criteria with their calculated photo-$z$ instead of the spec-z. By doing this, we retain most of our candidates with a similar fit stretch and, most importantly, we still reject all the CC-SNe as contaminants (even though their photo-$z$ was calculated with SN~Ia templates and can occasionally be very different from the spec-z). 

Perhaps the most concerning signature of contamination is the low-$s$ SN~Ia sample at $z>0.6$. These have rather blue colors and could be CC-SNe instead. As an additional test, we analyzed the host galaxy properties (Figure~\ref{hostprop}). In general these objects are found in massive and passive galaxies, not characteristic of CC-SNe. This provides evidence that most of these candidates, including the $z>0.6$ population, are likely the actual blue tail of the low-$s$ SN~Ia distribution, as mentioned earlier. 

Nevertheless, two of these objects, 05D3hp (with $s=0.59\pm0.04$ and $c=-0.16\pm0.09$) and 07D2bc (with $s=0.72\pm0.04$ and $c=-0.13\pm0.03$), have unusually blue colors at very low-stretch. The host of 05D3hp is small ($M=10^{7.9}M_{\odot}$) and star-forming ($\mathrm{SFR}/\mathrm{mass}=10^{-9.8}M_{\odot}\mathrm{yr}^{-1}$) and points rather towards a CC event imitating a low-$s$ SN~Ia. 07D2bc happened also in a star-forming ($\mathrm{SFR}/\mathrm{mass}=10^{-10.0}M_{\odot}\mathrm{yr}^{-1}$) but rather massive galaxy ($M=10^{11.0}M_{\odot}$). If these two objects are CC-events, they would constitute a $\sim5\%$ contamination fraction. Although it is dangerous to assume the local star formation from general galaxy properties, if we were to assume that all events in ``burst'' hosts with $\mathrm{SFR}/\mathrm{mass}<-9.5M_{\odot}\mathrm{yr}^{-1}$ are contaminants, then $\sim30\%$ of the $z>0.6$ objects are missclassified ($\sim25\%$ at $z\leq0.6$), in agreement with our generous estimates.

\section{Discussion}\label{disc}
We now turn to the interpretation of the low-$s$ SN~Ia rate in the light of suggested rate models, and compare it to different stretch populations. We investigate the empirical two-component rate evolution model of \citet{Mannucci05} and \citet{Scannapieco05} but also study simple delay-time distributions \citep{Dahlen99,Strolger04} for SN~Ia progenitors. A local low-$s$ rate estimate based on nearby studies helps constrain the models.

\subsection{Low-$z$ subluminous rate}
Recently, \citet{Li10c} have calculated a low-redshift SN~Ia rate and a corresponding $17.9^{+7.2}_{-6.2}\%$ fraction of 91bg-like objects \citep{Li10b}. To obtain a rough local rate estimate of low-$s$ SNe~Ia defined according to our study, i.e. $s\leq0.8$, we use their luminosity-function sample (Table~3 of \citealt{Li10b}) and directly fit objects for which we have photometric data (48 of 74) with SiFTO. For the rest we fit instead their closest $R$-band LC template match (column 11). Using the completeness-corrected numbers (column 9), we obtain 16.4 objects with $s\leq0.8$, or a $21.9^{+6.9}_{-5.4}\%$ fraction. This fraction is complete within a volume of 80 Mpc and translates to a low-$s$ rate of $(6.3\pm1.8)\times10^{-6}\mathrm{yr}^{-1}\mathrm{Mpc}^{-3}h_{70}^3$ based on the local rate of \citet{Li10c} (adjusted for $h_{70}=1$) and including the uncertainty from the Poisson error. We note that by using the whole low-$z$ training sample of \S\ref{lowz}, the 41 low-$s$ objects correspond to a 22\% fraction of the total, in good agreement.


This rate estimate is based on small sample statistics and suffers from the biases that arise from a distance-limited host-targeted survey, where the rate of SNe is converted to a volumetric rate via the galaxy luminosity function. This can lead to systematic errors, as brighter galaxies may contain more SNe~Ia of a certain type (as is expected for low-$s$). However, a local low-$s$ rate estimate is a valuable tool for comparison to the high-$z$ rates.

\subsection{Power-law evolution}
The low-$s$ SN~Ia rate, as opposed to the normal-$s$ population, does not seem to increase with redshift and follow the star formation history (SFH) trend found in P10 (Figure~\ref{rates}). If we fit a simple power-law, $r\propto(1+z)^{\alpha}$, to the SNLS rate to $z=0.6$ and the low-$z$ estimate, we obtain $\alpha=-0.90\pm1.38$, smaller than the value of the normal-$s$ population found by P10 ($\alpha_{\mathrm{normal}}=1.91\pm0.24$). This means that an evolution similar to the normal-$s$ population is rejected with $\sim94\%$ probability. If we include the corrected $z>0.6$ SNLS points, we find a higher slope more consistent with zero: $\alpha=-0.34\pm1.41$. 

\subsection{Delayed component model} \label{hosts}

\begin{figure*}[htbp]
  \centering
  \includegraphics[width=0.45\linewidth]{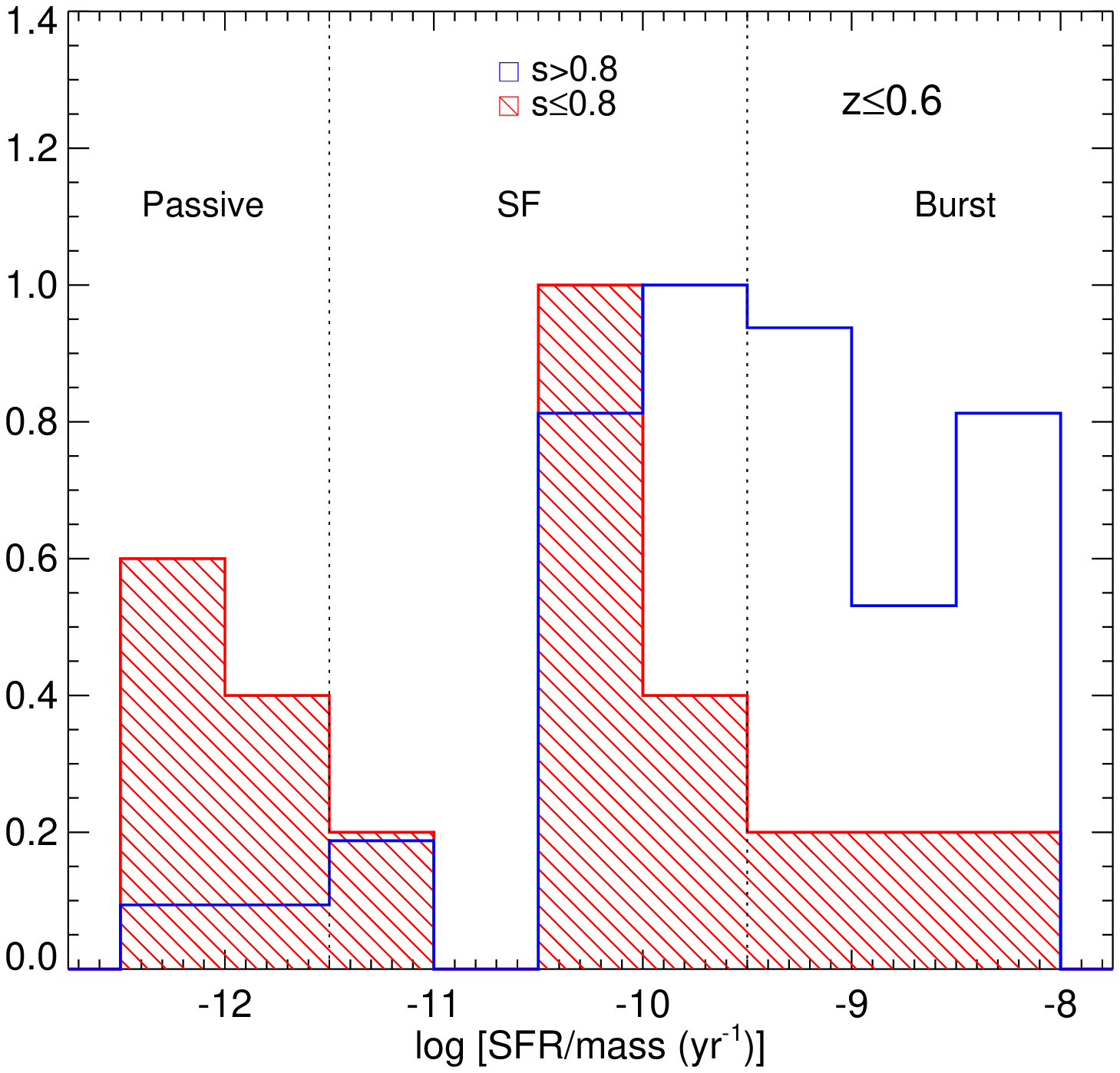}
  \includegraphics[width=0.445\linewidth]{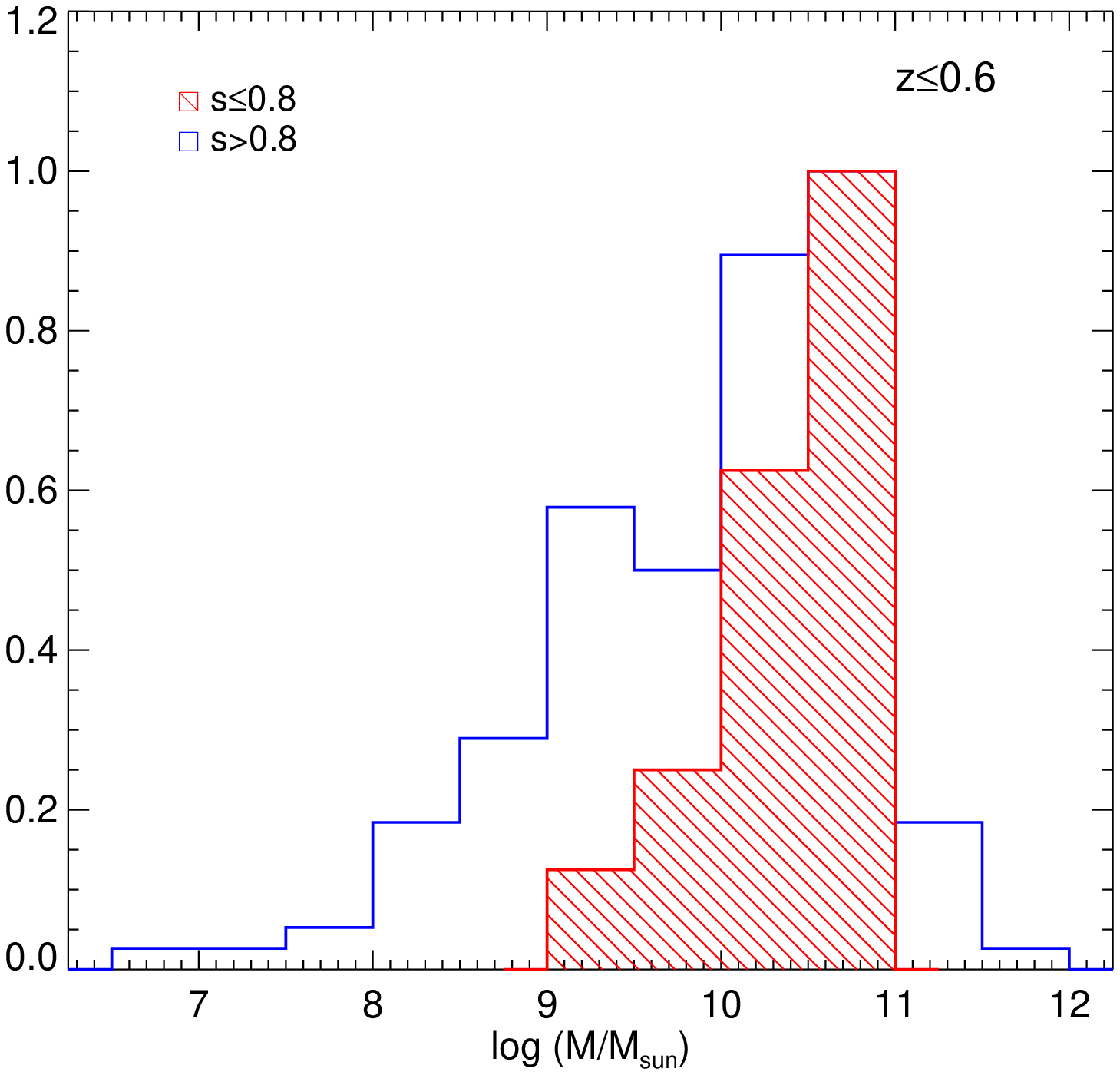}
  \includegraphics[width=0.45\linewidth]{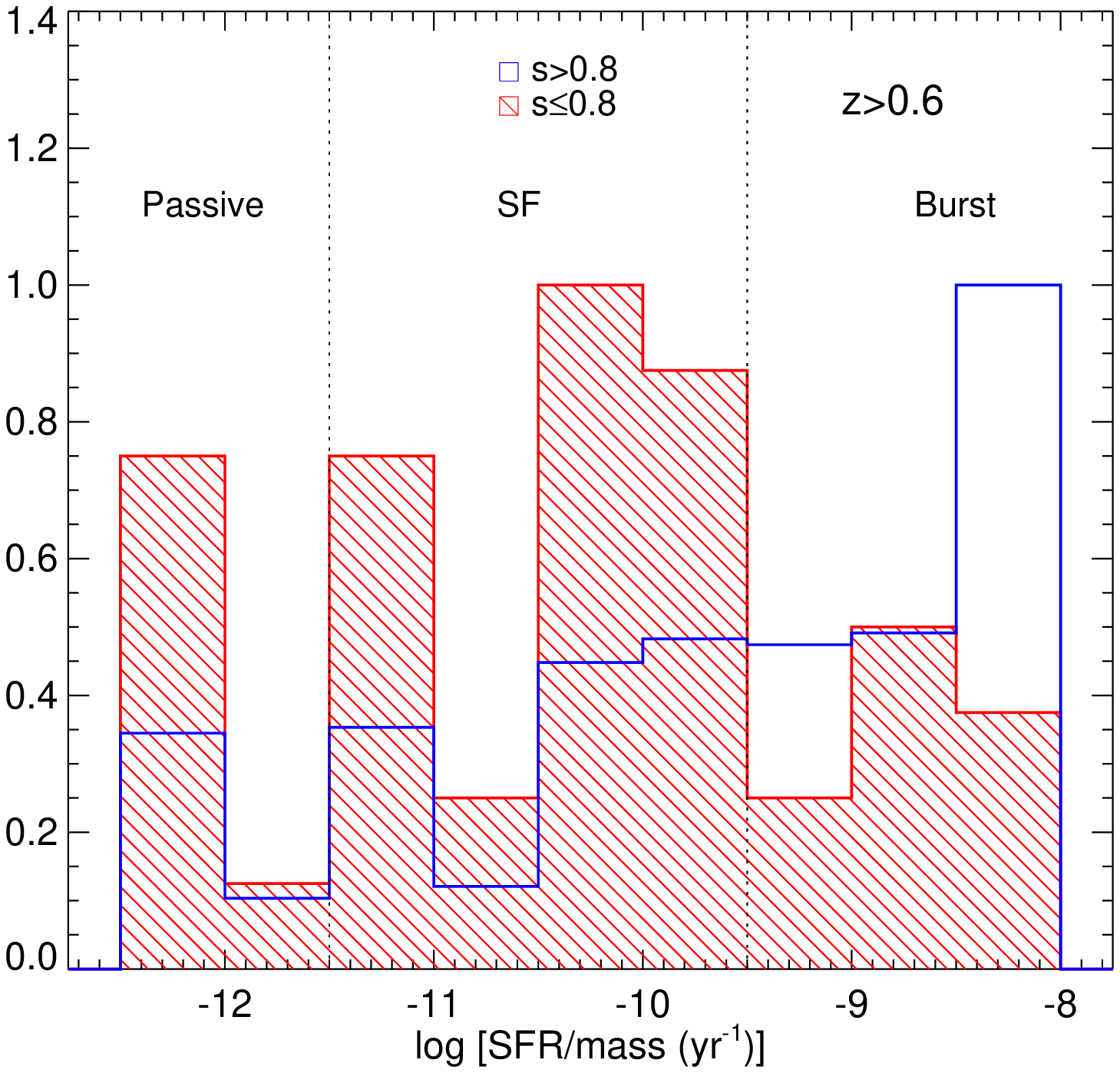}
  \includegraphics[width=0.445\linewidth]{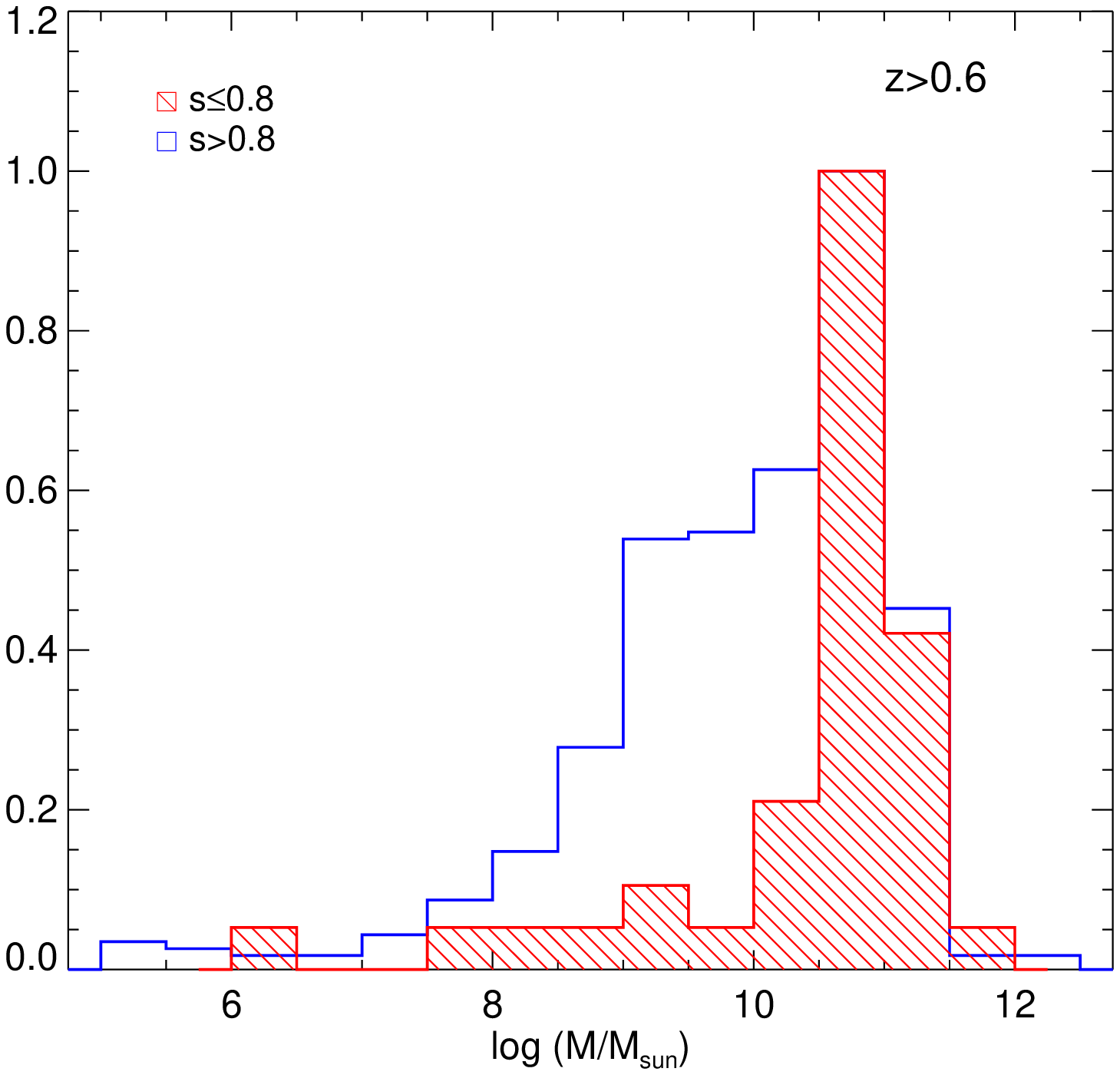}
		\caption{Host properties for $z<0.6$ (\emph{upper}) and $z>0.6$ (\emph{lower}) SN~Ia candidates. \emph{Left:} Normalized specific star formation (star formation per unit mass) distribution divided in normal-$s$ (blue) and low-$s$ (red) populations. The probability that the two populations are drawn from the same distribution (KS-statistic) is less than $\sim0.5\%$ for both redshift ranges. \emph{Right:} Normalized host stellar mass distribution per stretch range, normal-$s$ (blue), and low-$s$ (red). The KS-statistic is below $\sim7\%$ for both redshift ranges.}
        \label{hostprop}
\end{figure*}

It is known that SNe~Ia occur in both star forming and non-star forming regions, whereas low-$s$ objects are usually found in passive ones. In the model of \citet{Mannucci05,Scannapieco05}, two populations are considered: a ``prompt'' component for SNe exploding in 0.1Gyr or less that is proportional to the star formation history, $\dot{M}(z)$ and a ``delayed'' component for events with higher delay times that is proportional to the host mass, $M(z)$. It is described as:

\begin{equation}
SNR_{\mbox{Ia}}(z)=A M(z)+B\dot{M}(z),
\end{equation}

where A and B are scale factors for both components. The delayed model for low-$s$ SNe~Ia is supported when one looks at the host properties of the low-$s$ sample. CFHT Legacy Survey host galaxy broadband photometry can be fitted to several template galaxy SEDs using PEGASE/ZPEG \citep{LeBorgne02} to determine average galaxy star formation rates (SFRs), stellar masses and photometric redshifts \citep{Sullivan06b}. The properties of SNe~Ia are compared to their host type (see Figure~\ref{hostprop}) to investigate the behavior of low-$s$ objects in contrast to the normal-$s$ population.

We use the specific star formation rate (sSFR), the star formation per unit stellar mass, as the indicator of star formation activity, and separate the host galaxy sample in passive galaxies, $\log(\mathrm{sSFR}(\mathrm{yr}^{-1}))<-11.5$, weak star formers, $-11.5\leq\log(\mathrm{sSFR}(\mathrm{yr}^{-1}))\leq-9.5$, and strong star formers $\log(\mathrm{sSFR}(\mathrm{yr}^{-1}))>-9.5$, where SFR refers to the average star formation over the last 0.5~Gyr. This approach ignores important internal variations of SF in the galaxy, as shown in \citet{Raskin09b}, and provides only a general estimate of the SF. The sSFR distribution for two different stretch ranges (left Figure~\ref{hostprop}), normal-$s$ and low-$s$, provides clear evidence that the low-$s$ population is less dependent on star formation. \citet{Neill10} find that no local low-$s$ SN~Ia is found in a host with $\log(\mathrm{sSFR}(\mathrm{yr}^{-1}))>-9.5$. Here, we corroborate their results for high-$z$ objects finding that only 3 ($17\%$) at $z<0.6$, and 11 ($25\%$) at higher redshift, have higher mass-weighted star formation rates. 

Moreover, the host mass distribution for these stretch ranges (Figure~\ref{hostprop}, right) shows clearly different distributions, in which low-$s$ SNe~Ia occur almost solely in massive hosts. We extend the results of \citet{Neill10} to the distant universe: low-$s$ SNe~Ia are found mostly in hosts with masses above $\sim10^{9.5}M_{\odot}$. For all cases, the distribution of host-galaxy mass for the two populations, normal-$s$ and low-$s$, have a low two-sided Kolmogorov-Smirnov (KS) statistic, indicating a small probability that they are randomly drawn from the same distribution. 

The above relations of SN~Ia properties with host characteristics demonstrate that previous findings extend to the extreme of the SN~Ia population at $z>0.1$, and at the same time are a confirmation that the observed photometric low-$s$ sample obtained is real and not strongly contaminated from core-collapse events that are only found in late-type galaxies. 

They also provide additional support to model the low-$s$ SN~Ia rate as just an $A$-component. Fitting the latter to the local and SNLS and rates up to $z=0.6$ provides $A=(1.05\pm0.28)\times10^{-14}\,\mathrm{SNe}\,\mathrm{yr}^{-1}M_{\odot}^{-1}$ and is shown in Figure~\ref{rates}. If we fit both components, we obtain $A=(1.29\pm0.66)\times10^{-14}\,\mathrm{SNe}\,\mathrm{yr}^{-1}M_{\odot}^{-1}$ and $B=(-0.27\pm0.64)\times10^{-4}\,\mathrm{SNe}\,\mathrm{yr}^{-1}(M_{\odot}\mathrm{yr}^{-1})^{-1}$. When we include the corrected SNLS $z>0.6$ points, we obtain $A=(1.14\pm0.61)\times10^{-14}\,\mathrm{SNe}\,\mathrm{yr}^{-1}M_{\odot}^{-1}$ and $B=(0.03\pm0.62)\times10^{-4}\,\mathrm{SNe}\,\mathrm{yr}^{-1}(M_{\odot}\mathrm{yr}^{-1})^{-1}$. In the latter two cases, $B$ is close to zero ($B<0$ is an unphysical result) suggesting that the mass component dominates over the component proportional to the star formation history. 

Comparing our delayed component (for $z<0.6$) with the normal SN~Ia population, the ratio of this value to the one in P10 is $\frac{A_{\mathrm{low}-s}}{A_{\mathrm{P10}}}\simeq30\%$, which is particularly important at low-$z$, where the delayed component becomes dominant. This ratio sets a constraint on the fraction of low-$s$ progenitors.

\subsection{Delay-time distributions} \label{dtd}

Ideally, the rate should be able to help constrain the SN~Ia delay-time distribution as follows:

\begin{equation}\label{dtdeq}
\mbox{SNR}_{\mbox{Ia}}(t)=\nu \int_{t_F}^t \mbox{SFR}(t')\mbox{DTD}(t-t')dt',
\end{equation}
where $t'$ is the age of the universe, $t_F$ the time at which the first stars formed (effectively zero for the SNLS), DTD$(t_d)$ is the delay-time ($t_d$) distribution and $\nu$ is the number of SNe~Ia that explode per unit stellar mass formed. This number is related to the progenitor mass range and IMF, $\Psi(M)$, as well as the efficiency $\eta$ or fraction of white dwarfs that explode as SNe~Ia, and the range of WD progenitor masses, $M_{\mathrm{WD}}^{\mathrm{low}}$ and $M_{\mathrm{WD}}^{\mathrm{up}}$:

\begin{equation}\label{imf}
\nu=\eta\frac{\int_{M_{\mathrm{WD}}^{\mathrm{low}}}^{M_{\mathrm{WD}}^{\mathrm{up}}}\Psi(M)dM}{\int_{0.1M_{\odot}}^{125M_{\odot}}M\Psi(M)dM}
\end{equation}

Eq.~\ref{dtdeq} can set constraints on the progenitors through the delay-time distribution if the SFH is known. We assume here the parameterization of \citet{Hopkins06} for a SalA IMF (h=0.7):
\begin{equation}
 \mathrm{SFH}\propto\frac{(0.017+0.13z)h}{1+(z/3.3)^{5.3}}.
 \end{equation}

A simple parametrization power-law DTD $\propto t^n$, has been used by several authors to model the SN~Ia rate. $n\simeq-1$ is obtained in the SN~Ia rate studies made by \citet{Totani08,Maoz10c,Maoz10b,Maoz10a} and is consistent with the DD scenario. \citet{Pritchet08} obtain a different power-law of $n\simeq-0.5$ investigating the SN~Ia rate as a function of sSFR in the SNLS. It is interesting to see how the sub-sample of low-$s$ behaves, in particular as the inferred DTDs tend to have a low tail at higher delay-times (after a sharp initial peak) where the subluminous SNe~Ia are expected to lie.

\begin{figure}[htbp]
  \centering
  \includegraphics[width=1.0\linewidth]{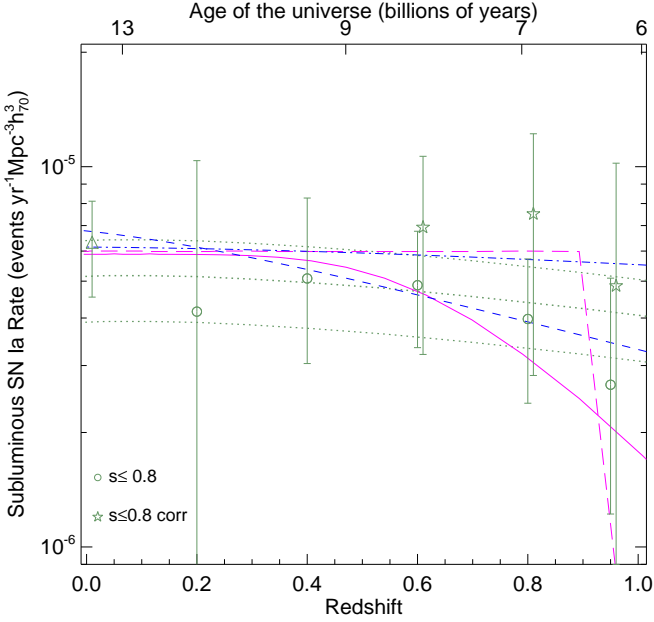}
    \caption{Best fit DTDs to the low-$s$ SN~Ia rate. Up to $z<0.6$: power-law (dashed blue) with $t^{-0.93\pm2.13}$, and Gaussian behaviour (solid purple) with $\tau=(5.11\pm5.65)$Gyr and $\sigma=0.94\pm3.16$. All corrected rates: power-law (dot-dashed blue) with $t^{-2.67\pm6.98}$ and Gaussian behaviour (long dashed purple) with $\tau=(5.88\pm4.84)$Gyr and $\sigma=0.15\pm2.41$. The dotted green lines represent the best $A$ model and errors for $z<0.6$ (see \S\ref{hosts}).}
        \label{rates_dtd}
\end{figure}

By fitting this model to the local and SNLS data up to $z=0.6$, we obtain $n=-0.93\pm2.13$ reflecting the large uncertainties in the rates. With the corrected $z>0.6$ rates, the power-law becomes $n=-2.67\pm6.98$ (see Figure~\ref{rates_dtd}). This is accomplished by increasing the other free parameter, the scale $\nu$, by an order of magnitude such that many more SNe~Ia happen at early times, in order to compensate for the steady rates at higher-$z$. 

We also use the simple empirical models employed in \citet{Strolger04}: an exponential function motivated by the double-degenerate scenario \citep{Tutukov94} and a Gaussian function motivated by a DTD with a specific delay-time \citep{Dahlen99}. The Gaussian distribution can be a proper approach for sub-groups of SNe~Ia, such as the low-$s$ one, but not for the entire population, which has been shown to be either bimodal or slowly declining in time. We use the formalism of the unimodal skew-normal DTD of eq.6 of \citet{Strolger10} as a generalization of these models encompassing a wide variety of shapes and distributions.


The model of \citet{Strolger10} has three independent variables, We obtain for the local and $z<0.6$ rates: $\xi=6.4\pm3.8$, $\omega=1.6\pm5.2$ and $\alpha=-81.2\pm26400$, which translate to a Gaussian behaviour of characteristic delay-time $\tau=(5.11\pm5.65)$ Gyr and width $\sigma=0.94\pm3.16$. This fit is shown in Figure~\ref{rates_dtd}. It indicates that low-$s$ SNe~Ia come from progenitors 3-10 Gyr old following a single gaussian-shaped distribution. Such delay-times correspond to masses of $1.5-2.5M_{\odot}$ using $t\sim m^{-2.5}$. This sustains the argument that low-$s$ SNe~Ia come exclusively from a delayed channel of SNe~Ia. Such estimates are higher than the lower delay-time limit obtained by \citet{Schawinski09} based on individual SN~Ia studies in early-type galaxies with optical and UV-data. On the other hand, they agree with \citet{Brandt10} who find that SDSS SNe Ia with $s<0.92$ have delay-times of $\gtrsim2.4$Gyr. Fitting all points, corrected SNLS and low-$z$, we obtain $\tau=(5.88\pm4.84)$ Gyr and $\sigma=0.15\pm2.41$ ($\xi=5.8\pm3.6$, $\omega=0.2\pm4.0$ and $\alpha=0.9\pm17300$) instead. This approximates to a much narrower Gaussian spanning 5.3-6.3 Gyr ($1.95-2.10M_{\odot}$), time at which all low-$s$ SNe~Ia are created. This explains the sharp drop in the modeled rate at $z\sim0.9$ seen in Figure~\ref{rates_dtd}. Nevertheless, due to the large error uncertainties, all these results are poorly constrained.

The explosion fraction is around $0.0001-0.0003$ low-$s$ SNe~Ia per $M_{\odot}$ for all models except for the power-law fit to all corrected data ($t^{-2.7}$) which is $0.001$ low-$s$ SNe~Ia per $M_{\odot}$. These fractions are lower than for the normal-$s$ population --generally higher than $\sim10^{-3}$ \citep[e.g.][]{Dahlen04,Maoz10b}. Both populations differ by an order of magnitude which can be due either to a) a lower conversion efficiency $\eta$ of WD into SN~Ia, or b) a lower progenitor mass range, taking into account that low mass stars are more frequent, or a combination of both.  If one assumes that the conversion efficiency is constant over mass --this can be the case for certain considerations of the SD and DD scenarios \citep{Pritchet08}--  the allowed mass range of low-$s$ SNe~Ia would need to be extremely narrow. With a SalA IMF, and assuming that all main-sequence stars with $0.8M_{\odot}<M<8M_{\odot}$ become WDs, the cutoff value between low-$s$ and normal-$s$ progenitors (from Eq.~\ref{imf}) would lie at $\sim1M_{\odot}$, an extremely narrow range of stars with too long delay-times ($>10$ Gyr) to be seen at $z\gtrsim0.3$. Changing the range of progenitors masses to $3M_{\odot}<M<8M_{\odot}$ \citep{Nomoto94}, the cutoff would need to be at $\sim3.2M_{\odot}$. On the other hand, using the lower WD mass limit of $1.95M_{\odot}$ and cutoff at $2.10M_{\odot}$, from the Gaussian distribution previously obtained, we also get a consistent explosion fraction ratio of low-$s$ to normal-$s$.


\subsection{Extremely low-$s$ SNe~Ia}
\begin{figure}[htbp]
  \centering
  \includegraphics[width=0.85\linewidth]{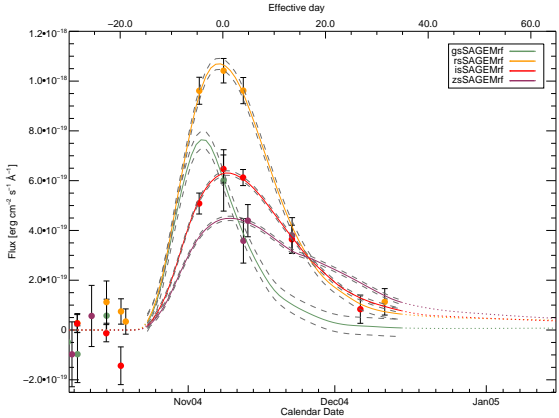}
		\caption{Extremely low-$s$ candidate, 04D4ms, at spectroscopic redshift $z=0.57$ with $s=0.63\pm0.08$ and $c=0.02\pm0.06$.}
        \label{vlows}
\end{figure}

We note that our sample does not consist of extremely subluminous SNe~Ia. Looking at the local sample of typical subluminous objects, such as SN1991bg or SN1999by, the stretch is on the low end, i.~e., $s=0.5-0.65$. In the SNLS $0.1<z<0.6$ range, we only obtain one such object (see Figure~\ref{vlows}). This SN has a spectroscopic redshift of $z=0.57$ with LC parameters: $s=0.63\pm0.08$ and $c=0.02\pm0.06$. The candidate is hosted in a galaxy of mass $M=10^{10.47}M_{\odot}$ and $\mathrm{sSFR}=10^{-11.93}M_{\odot}\mathrm{yr}^{-1}$, typical of very low-stretch objects. 

\begin{figure}[htbp]
  \centering
  \includegraphics[width=1.0\linewidth]{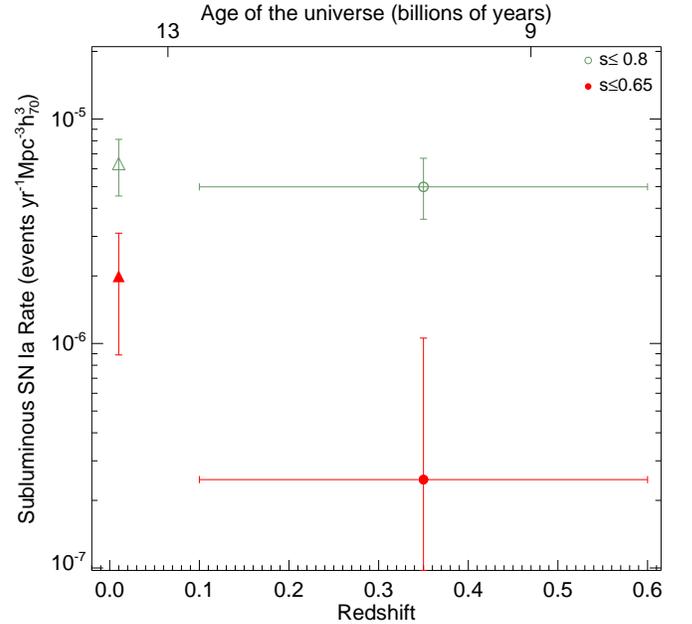}
		\caption{Single bin rate evolution for low-$s$ candidates ($s\leq0.8$) and extremely low-$s$ candidates with $s\leq0.65$ from the SNLS (green and red circles respectively), and the estimated fraction of the SN~Ia rate from \citet{Li10c} (triangles). Horizontal error bars indicate the width of the z-bin.}
        \label{vsub}
\end{figure}

This single SN corresponds to a single $z=0.35$ rate of $(0.25^{+0.56}_{-0.21}(\mathrm{stat})^{+0.06}_{-0.03}(\mathrm{MC}))\times10^{-6}\mathrm{yr}^{-1}\mathrm{Mpc}^{-3}$ (see Figure~\ref{vsub}). Despite the obvious uncertainty in this number, it is tempting to argue that perhaps this ``extremely low-$s$'' population, i.e. $s<0.65$, actually behaves differently than our $s\leq0.8$ group. If we maintain the extremely low-$s$ definition for the local luminosity-function sample of \citet{Li10b} as well, we obtain 5 objects with $s\leq0.65$ (5.16 corrected for completeness), which corresponds to $6.9^{+4.5}_{-2.9}\%$ of their rate (if we were to use the entire local sample, 15 objects have $s<0.65$, i.e., a $\sim9\%$ fraction), that is a rate of $(2.0\pm1.1)\times10^{-6}\mathrm{yr}^{-1}\mathrm{Mpc}^{-3}$. We note that this is much smaller than the 91bg fractions generally quoted, such as the $17.9^{+7.2}_{-6.2}\%$ by \citet{Li10b}, indicating that this photometric definition is more restrictive than their definition (which is also based on spectra). Contrarily, if we were to use their fraction of ``subluminous'' SNe~Ia to estimate the stretch limit, we would obtain $s\sim0.75$. We emphasize therefore the importance of having a clear and consistent definition of ``subluminous'' SNe~Ia across the redshift range. As shown in Figure~\ref{vsub}, the rate evolution for this population also reveals a decline, perhaps even more rapid than for our standard low-$s$ sample.

\section{Summary}\label{conclusions}

We have conducted a search in the SNLS for low-stretch SNe~Ia, defined here as objects with $s\leq0.8$. Typical ``subluminous'' surveys considered more extreme objects based also on spectroscopic features which have equivalent stretches of $\lesssim0.72-0.75$. The definition chosen suits better our constructed color-stretch and magnitude-color-stretch relations trained with low-$z$ and SNLS objects. A lower transition value, nevertheless, provides also consistent LC relations. This argues for a continuous distribution of LC properties of SNe~Ia, with low-$s$ SNe~Ia having distinctive redder colors and a steeper color-stretch relation. The exact stretch value at which this transition occurs is not a sharp one and lies somewhere between and $0.69-0.87$.

The LC relations serve for the typing of low-$s$ SNe~Ia at high redshift and for modeling the survey efficiency. We find 18 photometrically-identified low-stretch SNe~Ia up to $z\sim0.6$, indicating that the primary reason for their previous non-detections at high redshifts was due to selection effects. Most of the candidates at $z<0.6$ have $s>0.65$ and colors similar to the local sample. Only one of them is comparable to the faintest local counterparts such as 1991bg. The candidates are observed to be primarily hosted in passive galaxies, with no star formation, as it is found locally. The spectroscopic candidates have no conclusive features due to low signal-to-noise. 

We also find a population of blue low-$s$ SNe~Ia at $z>0.6$ hosted mostly in quiet host envirnoments. They represent the blue end of the low-$s$ SN~Ia sample at high-redshift, whereas its reddest counterpart is not seen due to selection effects. A $30\%$ fraction of these have unusual blue colors that are mostly attributed to the larger photometric uncertainties at high-$z$. Nonetheless, some of them occur in star-forming galaxies and are possible CC-SN contaminants ($\lesssim10\%$). Alternatively we speculate that they could represent a different set of objects that evolve rapidly but have bluer colors than typical low-$s$ SNe.

We calculate a rate evolution of low-$s$ SNe~Ia as a function of redshift and the results indicate a steady or slightly decreasing trend up to $z=0.6$. We correct the rates at $z>0.6$, where our sample is incomplete, and find the evolution to be different from the increase of the normal SN~Ia population. The missclassification of CC-SNe as SNe~Ia (or different unknown blue high-$z$ transients) is an important source of systematics in the rate measurements. This would overestimate the rate and should become larger with redshift, so that the effect of a decreasing rate would be even stronger. The main findings of this study are therefore not influenced by contamination.

An A-component model, proportional to just the mass of the galaxy, is sufficient to fit the rate. This is expected as most of the candidates are hosted in passive galaxies with no star formation. Low-$s$ SNe~Ia are therefore a population coming from evolved and/or metal-rich progenitors not dependent on the SFH. As shown in \citet{Howell07}, this has consequences for cosmological studies because the relative proportion of low to high stretch SNe~Ia evolves at high redshift. The present study supports their findings at the extreme low-stretch end of the population. We find that $A+B$ models have an $A$-component with $\sim30\%$ coming from the low-$s$ population.

Studies of the DTD of different scenarios help constrain progenitors masses and models, as well as the conversion efficiency of WDs into SNe~Ia. Despite the large uncertainties, the model of \citet{Strolger10} fitted to our data hints to a Gaussian distribution centered around $\tau\simeq5-6$ Gyr. The explosion fraction for those models is $\nu\sim10^{-3}-10^{-4}$ low-$s$ SNe~Ia per $M_{\odot}$, an order of magnitude lower than for normal-$s$ objects. These results again argue for the low-$s$ being part of a delayed component. Low-$s$ SNe~Ia, together with other long delay SNe~Ia of higher-$s$, could come from a separate progenitor channel. Although both, the SD and DD, scenarios can produce delayed SNe~Ia, the merger of two low-mass WDs producing a sub-Chandrasekhar explosion seems more favorable to produce faint SNe in old stellar environments.

Finally, future surveys should be able to determine the rate evolution of low-$s$ much better and to higher redshifts where the systematic errors in this study become more significant. Spectroscopic redshifts of high-$z$ low-$s$ SNe~Ia are essential for this and could provide additional hint on their properties and their evolution. Further comparison of different SN~Ia populations will give clues to understand the formation and explosion mechanisms. 

\vspace{20mm}
We are grateful to the CFHT Queued Service Observations Team and the entire SNLS collaboration. This work has used observations obtained with MegaPrime/MegaCam, a joint project of CFHT and CEA/DAPNIA, at the Canada-France-Hawaii Telescope (CFHT) which is operated by the National Research Council (NRC) of Canada, the Institut National des Sciences de l'Univers of the Centre National de la Recherche Scientifique (CNRS) of France, and the University of Hawaii. This work is also based in part on data products produced at the Canadian Astronomy Data Centre as part of the CFHT Legacy Survey, a collaborative project of NRC and CNRS. We also use observations taken at the Gemini Observatory, which is operated by the Association of Universities for Research in Astronomy, Inc., under a cooperative agreement with the NSF on behalf of the Gemini partnership: the National Science Foundation (United States), the Science and Technology Facilities Council (United Kingdom), the National Research Council (Canada), CONICYT (Chile), the Australian Research Council (Australia), Minist\'erio da Ci\^encia e Tecnologia (Brazil) and Ministerio de Ciencia, Tecnolog\'ia e Innovaci'on Productiva (Argentina). The Gemini program identification numbers are: GS-2003B-Q-8, GN-2003B-Q-9, GS-2004A-Q-11, GN-2004A-Q-19, GS-2004B-Q-31,GN-2004B-Q-16, GS-2005A-Q-11, GN-2005A-Q-11, GS-2005B-Q-6, GN-2005B-Q-7, GN-2006A-Q-7 and GN-2006B-Q-10. We acknowledge the support from our funding agencies: NSERC, CIAR, CNRS and CEA. MS acknowledges support from the Royal Society.

\bibliographystyle{apj}
\bibliography{astro}

\begin{table}[h!t]
 \centering
 \caption{Nearby low-$s$ objects that passed the observation and fit culls used in the analysis}
  \begin{tabular}{|c|c|c|c|}
    \hline
    \hline
    Name & Redshift & Stretch & Source \\
    \hline
  SN1986g &  0.002 & 0.71 &  \citet{Phillips87}   \\
  SN1990af &  0.050 &0.75 &  \citet{Hamuy96}   \\
  SN1991bg & 0.004 & 0.50  &  \citet{Turatto96},\citet{Filippenko92b}   \\
  SN1992bl & 0.044 & 0.80 & \citet{Hamuy96}  \\
  SN1992bo & 0.019 & 0.79 & \citet{Hamuy96}   \\
  SN1992br & 0.088 & 0.60 & \citet{Hamuy96}   \\
  SN1993h  & 0.024 & 0.66 & \citet{Hamuy96},\citet{Altavilla04}  \\ 
  SN1994m  & 0.023 & 0.76 & \citet{Riess99},\citet{Altavilla04}  \\
  SN1995ak &  0.023 & 0.79 & \citet{Riess99} \\
  SN1998bp &  0.010 & 0.63 & \citet{Jha06}   \\
  SN1998co &  0.018 & 0.65 & \citet{Jha06}   \\ 
  SN1998de &  0.017 & 0.64 & \citet{Modjaz01}   \\ 
  SN1999bm &  0.143 & 0.68 & \citet{Kowalski08}   \\
  SN1999by &  0.002 & 0.61 & \citet{Garnavich04}   \\
  SN1999da &  0.013 & 0.59 & \citet{Krisciunas01}    \\
  SN2000dk &  0.017 & 0.77 & \citet{Jha06}  \\
  SN2001da &  0.017 & 0.61 & \citet{Hicken09a}  \\
  SN2002cx &  0.024 & 0.57 & \citet{Phillips07}   \\
  SN2002hw &  0.018 & 0.73 & \citet{Hicken09a}  \\
  SN2003ic &  0.056 & 0.77 & \citet{Hicken09a}  \\ 
  SN2003iv &  0.034 & 0.74 & \citet{Hicken09a}  \\
  SN2005am &  0.008 & 0.71 & \citet{Hicken09a},\citet{Li06}   \\
  SN2005bl &  0.024 & 0.60 & \citet{Taubenberger08} \\ 
  SN2005hf &  0.043 & 0.76 & \citet{Hicken09a}  \\ 
  SN2005ke &  0.005 & 0.69 & \citet{Hicken09a}  \\
  SN2005mc &  0.025 & 0.69 & \citet{Hicken09a}  \\
  SN2006ak &  0.038 & 0.78 & \citet{Hicken09a}  \\
  SN2006al &  0.068 & 0.78 & \citet{Hicken09a}  \\
  SN2006br &  0.025 & 0.78 & \citet{Hicken09a}  \\
  SN2006bw &  0.030 & 0.71 & \citet{Hicken09a}  \\ 
  SN2006bz &  0.028 & 0.53 & \citet{Hicken09a}  \\
  SN2006gj &  0.028 & 0.71 & \citet{Hicken09a}  \\
  SN2006gt &  0.045 & 0.53 & \citet{Hicken09a}  \\
  SN2006hb &  0.015 & 0.69 & \citet{Hicken09a} \\ 
  SN2006je & 0.038 & 0.48  & \citet{Hicken09a} \\
  SN2006mo &  0.037 & 0.73 & \citet{Hicken09a}  \\
  SN2006n &  0.014 & 0.77 & \citet{Hicken09a}  \\ 
  SN2006nz &  0.038 & 0.62 & \citet{Hicken09a}  \\
  SN2006ob &  0.059 & 0.75 & \citet{Hicken09a}  \\
  SN2006td & 0.016 &  0.79 & \citet{Hicken09a}  \\ 
  SN2007al & 0.012 & 0.47 & \citet{Hicken09a}   \\ 
  SN2007au &  0.021 & 0.67 & \citet{Hicken09a}  \\
  SN2007ax &  0.007 & 0.53 & \citet{Hicken09a}  \\
  SN2007ci &  0.018 & 0.79 & \citet{Hicken09a}  \\
\hline
  \end{tabular}
  \label{table_lows_objects}
\end{table}

\begin{table}[h!t]
 \centering
\caption{Nearby non-Ia SNe with acceptable SN~Ia SiFTO fits, enough data coverage and fitted $s\leq0.8$.}
  \begin{tabular}{|c|c|c|c|c|}
    \hline
    \hline
    Name & Type & Redshift & Stretch & Source \\
    \hline
SN1999eu & SNIIP & 0.004 & 0.45  & \citet{Pastorello04} \\
SN2001b & SNIb   &  0.005 & 0.60   &   \citet{Tsvetkov06a}   \\
SN2002ao & SNIbn  &   0.005 & 0.54   &  \citet{Pastorello08}   \\
SN2002ap & SNIc  &   0.002 &  0.68   &  \citet{GalYam02,Foley03,Yoshii03}   \\
SN2003jd & SNIc  &  0.019 &   0.73   &   \citet{Valenti08}   \\
SN2006aj & SNIc  &  0.033 &  0.73   &   \citet{Pian06,Sollerman06}   \\
\hline
  \end{tabular}
 \label{table_cont}
\end{table}


\begin{table}[h!t]
 \centering
   \caption{Culls applied to the training sample} 
 \begin{tabular}{|c|c|c|c|c|}
    \hline
    \textbf{Cull} & \multicolumn{4}{|c|}{\textbf{Number}} \\ 
         & \multicolumn{2}{|c|}{confirmed} & \multicolumn{2}{|c|}{confirmed} \\
         & \multicolumn{2}{|c|}{SNe~Ia(low-$s$)}  & \multicolumn{2}{|c|}{CC-SNe(low-$s$)} \\
         & low-$z$ & SNLS & low-$z$ & SNLS \\ 
    \hline
     None                               & 246(57) & 112(3) & 37(12) & 47(9) \\
     observation                        & 207(44) & 106(3) & 14(6) & 21(4)\\
     $\chi^2_{\mathrm{SiFTO}}$ and snakes  & 184(41) & 97(3) & 2(2) & 1(0)\\
\hline
  \end{tabular}
\label{table_train_culls}
\end{table}


\begin{table}[h!t]
 \centering
   \caption{Culls applied to the SNLS sample} 
 \begin{tabular}{|c|c|c|c|c|}
    \hline
    \textbf{Cull} & \multicolumn{4}{|c|}{\textbf{Number}} \\ 
         & all & low-$s$ & confirmed & confirmed \\
         & & & SNe~Ia  & CC-SNe \\  
    \hline
     None                & 5234 & 1497 & 407  &  86    \\
     observation         & 2615 &  837 & 343  &  62    \\
     $\chi^2_{\mathrm{SiFTO}}$     & 1054 &  144 & 338  &  22    \\
     $\chi^2_{\mathrm{estimate\_sn}}$ &  925 &  118 & 329  &   8    \\
     $\chi^2_{f , \mathrm{estimate\_sn}}$        &  848 &  109 & 303  &   5    \\
     snakes              &  710 &   70 & 295  &   1    \\
    \hline
     FINAL               &  710 &   70 & 295  &   1    \\
     $z<0.6$               &  174 &   18 & 118  &   0    \\
\hline
  \end{tabular}
\label{table_culls}
\end{table}

\begin{table}[h!t]
 \centering
\caption{Volumetric rates for the subluminous SNe~Ia in the SNLS\\
First and second column show the redshift at the middle of the bin and the mean redshift respectively. The third column is the number of objects per $z$-bin and the fourth is the mean weight ($w_i=(1+z_i)/(\epsilon_i\times\Delta T_i)$, where $\epsilon_i$ is the efficiency) used for the Poisson errors.  Fifth column indicate the measured rate with statistical weighted Poisson errors, and sixth column the Monte-Carlo rates with weighted Poisson, Monte-Carlo and contamination errors. The seventh column shows the correction factor (with errors) applied to get the last column high-$z$ rates corrected for the color distribution at $z<0.6$.}
  \begin{tabular}{|c|c|c|c|c|c|c|c|}
    \hline
    \hline
    $z_{\mathrm{mid}}$ & \textbf{$<z>$} & \textbf{$N$} & \textbf{$<w>$,\,$\sigma_w$} & \textbf{$r_{\mathrm{meas}}$} & \textbf{$r_{\mathrm{MC}}$}   & $f_{\mathrm{corr}}$ & \textbf{$r_{\mathrm{corr}}$}\\
                     &       &     &    & $(10^{-5}\mathrm{yr}^{-1}\mathrm{Mpc}^{-3}h_{70}^3)$ & $(10^{-5}\mathrm{yr}^{-1}\mathrm{Mpc}^{-3}h_{70}^3)$ & & $(10^{-5}\mathrm{yr}^{-1}\mathrm{Mpc}^{-3}h_{70}^3)$ \\
    \hline
    0.20 & 0.17 & 2 & $1.40,\,0.75$ & $0.46^{+0.68}_{-0.31}$ & $0.42^{+0.61}_{-0.28}(\mathrm{stat})^{+0.01}_{-0.17}(\mathrm{MC})\pm0.08(\mathrm{cont})$ & --& -- \\
    0.40 & 0.43 & 9 & $1.08,\,0.24$ & $0.74^{+0.41}_{-0.28}$ & $0.51^{+0.30}_{-0.21}(\mathrm{stat})^{+0.16}_{-0.08}(\mathrm{MC})\pm0.11(\mathrm{cont})$ & --& --\\
    0.60 & 0.63 & 19 & $1.16,\,0.21$ & $0.65^{+0.19}_{-0.15}$ & $0.49^{+0.20}_{-0.16}(\mathrm{stat})^{+0.12}_{-0.10}(\mathrm{MC})\pm0.13(\mathrm{cont})$ & $1.42\pm0.20$ & $0.69^{+0.35}_{-0.29}$\\
    0.80 & 0.78 & 18 & $1.44,\,0.65$ & $0.44^{+0.15}_{-0.11}$ & $0.40^{+0.24}_{-0.19}(\mathrm{stat})^{+0.11}_{-0.18}(\mathrm{MC})\pm0.12(\mathrm{cont})$ & $1.89\pm0.26$ & $0.75^{+0.51}_{-0.43}$ \\
    0.95 & 0.94 & 11 & $2.37,\,3.02$ & $0.48^{+0.30}_{-0.20}$ & $0.27^{+0.29}_{-0.19}(\mathrm{stat})^{+0.18}_{-0.18}(\mathrm{MC})\pm0.08(\mathrm{cont})$ & $1.82\pm0.40$ & $0.49^{+0.62}_{-0.40}$ \\
  \hline
\hline
  \end{tabular}
 \label{table_rates}
\end{table}

\end{document}